\documentclass[10pt]{elsarticle}

\usepackage{calrsfs,amsmath,amssymb,graphicx,array,accents,citesort,natbib,figlatex,epsfig}
\usepackage{setspace,wrapfig,marvosym, bbm, stmaryrd,centernot}

\def\Xint#1{\mathchoice
{\XXint\displaystyle\textstyle{#1}}%
{\XXint\textstyle\scriptstyle{#1}}%
{\XXint\scriptstyle\scriptscriptstyle{#1}}%
{\XXint\scriptscriptstyle\scriptscriptstyle{#1}}%
\!\int}
\def\XXint#1#2#3{{\setbox0=\hbox{$#1{#2#3}{\int}$}
\vcenter{\hbox{$#2#3$}}\kern-.5\wd0}}

\def\dashint{\Xint-}

\newcommand{\hs}{\hspace{0.01 in}}

\newlength{\kaka}

\newcommand{\ahref}[2]{}

\newcommand{\Zsup}{^{\text{\tiny o}}}

\newcommand{\obs}{^{\text{obs}}}

\newcommand{\beq}{\begin{equation}}
\newcommand{\eeq}{\end{equation}}
\newcommand{\lb}{\label}

\newcommand{\bea}{\begin{eqnarray}}
\newcommand{\eea}{\end{eqnarray}}
\newcommand{\bxr}{\begin{array}}
\newcommand{\exr}{\end{array}}

\newcommand\exs{\hspace*{0.4mm}}

\newcommand\nes{\hspace*{-0.4mm}}
\newcommand\nxs{\hspace*{-0.2mm}}

\newcommand{\bA} {\boldsymbol{A}}
\newcommand{\bU} {\boldsymbol{U}}
\newcommand{\bSig} {\boldsymbol{\Sigma}}
\newcommand{\bW} {\boldsymbol{W}}
\newcommand{\bC} {\boldsymbol{C}}
\newcommand{\bK} {\boldsymbol{K}}
\newcommand{\bV} {\boldsymbol{V}}

\newcommand{\bn} {\boldsymbol{n}}
\newcommand{\ba} {\boldsymbol{a}}
\newcommand{\bb} {\boldsymbol{b}}

\newcommand{\bx} {\boldsymbol{x}}
\newcommand{\by} {\boldsymbol{y}}
\newcommand{\be} {\boldsymbol{e}}
\newcommand{\bz} {\boldsymbol{z}}

\newcommand{\bd} {\boldsymbol{d}}
\newcommand{\bI} {\boldsymbol{I}}

\newcommand{\pff}{\boldsymbol{u}^{\textit{i}}}

\newcommand{\bksi} {\boldsymbol{\xi}}

\newcommand{\bfsig}{\boldsymbol{\sigma}}

\newcommand{\Bcal}{\mathrsfs{B}}
\newcommand{\Jcal}{J}

\newcommand{\Tcal}{{\sf T}}

\newcommand{\sip} {\!\cdot\!}

\newcommand{\So}{S\obs}

\newcommand{\G}{\Gamma}

\newcommand{\Bo}{\mathcal{B}_2}

\newcommand{\bzero}{\boldsymbol{0}}

\newcommand{\bu} {\boldsymbol{u}}
\newcommand{\tbu} {\tilde{\boldsymbol{u}}}
\newcommand{\bt} {{\boldsymbol{t}}}

\newcommand{\bbu} {\llbracket {\tbu}\rrbracket}
\newcommand{\bv} {\boldsymbol{v}}
\newcommand{\Ga}{\G_{\mbox{\tiny{trial}}}}
\newcommand{\bxi} {\boldsymbol{\xi}}
\newcommand{\bxio} {\bxi\Zsup}
\newcommand{\tbv} {\tilde{\boldsymbol{v}}}
\newcommand{\Geps}{\Gamma_{\!\varepsilon}}
\newcommand{\bbb} {\llbracket\bar{\boldsymbol{v}}\rrbracket}
\newcommand{\bbv} {\llbracket {\tbv}\rrbracket}

\newcommand{\dbbV} {\llbracket {\boldsymbol{V}}\rrbracket}

\begin{document}

\begin{frontmatter}

\title{On the elastic-wave imaging and characterization of fractures with specific stiffness}

\author[UMN]{Fatemeh Pourahmadian}
\author[UMN]{Bojan B. Guzina\corref{cor1}}

\address[UMN]{Department of Civil, Environmental \& Geo- Engineering, University of Minnesota, Twin Cities, MN 55455, USA}
\cortext[cor1]{Corresponding author. Tel. +612-626-0789, e-mail guzin001@umn.edu}

\begin{abstract} The concept of topological sensitivity (TS) is extended to enable simultaneous 3D reconstruction of fractures with unknown boundary condition and characterization of their interface by way of elastic waves. Interactions between the two surfaces of a fracture, due to e.g.~presence of asperities, fluid, or proppant, are described via the Schoenberg's linear slip model. The proposed TS sensing platform is formulated in the frequency domain, and entails point-wise interrogation of the subsurface volume by infinitesimal fissures endowed with interfacial stiffness. For completeness, the featured elastic polarization tensor - central to the TS formula -- is mathematically described in terms of the shear and normal specific stiffness $(\kappa_s, \kappa_n)$ of a vanishing fracture. Simulations demonstrate that, irrespective of the contact condition between the faces of a hidden fracture, the TS (used as a waveform imaging tool) is capable of reconstructing its geometry and identifying the normal vector to the fracture surface without iterations. On the basis of such geometrical information, it is further shown via asymptotic analysis -- assuming ``low frequency'' elastic-wave illumination,  that by certain choices of $(\kappa_s, \kappa_n)$ characterizing the trial (infinitesimal) fracture, the ratio between the shear and normal specific stiffness along the surface of a nearly-panar (finite) fracture can be qualitatively identified. This, in turn, provides a valuable insight into the interfacial condition of a fracture at virtually no surcharge -- beyond the computational effort required for its imaging. The proposed developments are integrated into a computational platform based on a 
regularized boundary integral equation (BIE) method for 3D elastodynamics, and illustrated via a set of numerical experiments.
\end{abstract} 


\end{frontmatter}

\newpage 

\section{Introduction} \label{sec1}

\noindent To date, inverse obstacle scattering remains a vibrant subject of interdisciplinary research with applications to many areas of science and engineering~\cite{Pik2002}. Its purpose is to recover the geometric as well as physical properties of unknown heterogeneities embedded in a medium from the remote observations of thereby scattered waveforms. Such goal is pursued by studying the nonlinear and possibly non-unique relationship between the scattered field produced by a hidden object, e.g.~fracture, and its characteristics.

From the mathematical viewpoint, the fracture reconstruction problem was initiated in~\cite{Kress1995} where, from the knowledge of the far-field scattered waveforms, the shape of an open arc was identified via the Newton's method. This work was followed by a suite of \emph{non-iterative} reconstruction approaches such as the factorization method~\cite[e.g.][]{Boukari2013}, the linear sampling method (LSM)~\cite{Kirsch2000} and the concept of topological sensitivity (TS)~\cite{Guzi2004} that are capable of retrieving the shape, location, and the size of buried fractures. Recently, a TS-related approach has also been proposed for the reconstruction of a collection of small cracks in elasticity~\cite{Ammari2013}.

A non-iterative approach to inverse scattering which motivates the present study is that of TS~\cite{Guzi2004,BonnetGuzina}. In short, the TS quantifies the leading-order perturbation of a given misfit functional due to the nucleation of an infinitesimal scatterer at a sampling point in the reference (say intact) domain. The resulting TS distribution is then used as \emph{anomaly indicator} by equating the support of its most pronounced negative values with that of a hidden  scatterer. The strength of the method lies in providing a computationally efficient way of reconstructing distinct inner heterogeneities without the need for prior information on their geometry. Recently,~\cite{Bellis2013} demonstrated the ability of TS to image \emph{traction-free cracks}. Motivated by the reported capability of TS to not only image -- but also characterize -- elastic inclusions~\cite{Ivan2007}, this study aims to explore the potential of TS for simultaneous imaging and interfacial characterization of fractures with contact condition due to e.g.~the presence of asperities, fluid, or proppant at their interface. 

Existing studies on the sensing of obstacles with unknown contact condition reflect two principal concerns, namely: i)~the effect of such lack of information on the quality of geometric reconstruction, and ii)~the retrieval -- preferably in a non-iterative way -- of the key physical characteristics of such contact. The former aspect is of paramount importance in imaging stress corrosion fractures~\cite{Hernandez2014}, where the crack extent may be underestimated due to interactions at its interface, leading to a catastrophic failure. To help address such problem, \cite{Boukari2013} developed the factorization method for the shape reconstruction of acoustic \emph{impedance} cracks. Studies deploying the LSM as the reconstruction tool~\cite[e.g.][]{Fiora2014,Fiora2010}, on the other hand, show that the LSM is successful in imaging obstacles and fractures \emph{regardless} of their boundary condition. As to the second concern, a variational method was proposed in~\cite{Colton2004} to determine the essential supremum of electrical impedance at the boundary of partially-coated obstacles. By building on this approach, \cite{Fiora2013} devised an iterative algorithm for the identification of surface properties of obstacles from acoustic and electromagnetic data. Recently, \cite{Minato2014} proposed a Fourier-based algorithm using reverse-time migration and wavefield extrapolation to retrieve the location, dip and heterogeneous compliance of an elastic interface under the premise of a)~one-way seismic wavefield, and b)~absence of evanescent waves along the interface.              

Considering the small-amplitude elastic waves that are typically used for seismic imaging and non-destructive material evaluation, the Schoenberg's \emph{linear slip model}~\cite{Schoenberg1980} is widely considered as an adequate tool to describe the contact condition between the faces of a fracture. This framework can be interpreted as a linearization of the interfacial behavior about the elastostatic equilibrium state~\cite{Fatemeh2012} prior to elastic-wave excitation, which gives rise to linear (normal and shear) specific stiffnesses $k_n$ and~$k_s$. Here it is worth noting that strong correlations are reported in the literature~\cite{Verdon2013, pyrak2014,Fatemeh2010} between $(k_s,k_n)$ and surface roughness, residual stress, fluid viscosity (if present at the interface), intact material properties, fracture connectivity, and excitation frequency. In this vein, remote sensing of the specific stiffness ratio~$k_s/k_n$ has recently come under the spotlight in hydraulic fracturing, petroleum migration, and Earth's Critical Zone studies~\cite{Georev2010,Verdon2013(2)}. By way of laboratory experiments~\cite{pyrak2014, Place2014, Bakulin2000}, it is specifically shown that~$k_s/k_n$ -- often approximated as either one (dry contact) or zero (isolated fluid-filled fracture) -- can deviate significantly from such canonical estimates, having fundamental ramifications on the analysis of the effective moduli and wave propagation in fractured media. A recent study~\cite{Verdon2013(2),Verdon2013} on the production from the Cotton Valley tight gas reservoir, using shear-wave splitting data, further highlights the importance of monitoring $k_s/k_n$ during hydraulic fracturing via the observations that:~i)~the correlation between proppant introduction and dramatic increase in $k_s/k_n$ can be used as a tool to directly image the proppant injection process; ii)~the ratio $k_s/k_n$ provides a means to discriminate between newly created, old mineralized and proppant-filled fractures, and~iii)~$k_s/k_n$ may be used to monitor the evolving hydraulic conductivity of an induced fracture network and subsequently assess the success of drilling and stimulation strategies.

In what follows, the TS sensing platform is developed for the inverse scattering of time-harmonic elastic waves by fractures with unknown geometry and contact condition  in~$\mathbb{R}^3$. On postulating the nucleation of an infinitesimal penny-shaped fracture with constant (normal and shear) interfacial stiffnesses at a sampling point, the TS formula and affiliated elastic polarization tensor are calculated and expressed in closed form. Simulations demonstrate that, irrespective of the contact condition between the faces of a hidden fracture, the TS is capable of reconstructing its geometry and identifying the normal vector to the fracture surface without iterations. Assuming illumination by long wavelengths, it is further shown that the TS is capable (with only a minimal amount of additional computation) of qualitatively characterizing the ratio $k_s/k_n$ along the surface of nearly-planar fractures. The proposed developments are integrated into a computational platform based on a regularized boundary integral equation (BIE) method for 3D elastodynamics. For completeness, the simulations also include preliminary results on the ``high''-frequency TS sensing of fractures with specific stiffness, which may motivate further studies in this direction. 

\section{Preliminaries} \label{sec2} 

\noindent Consider the scattering of time-harmonic elastic waves by a smooth fracture surface $\Gamma \subset \Bcal_1 \subset \mathbb{R}^3 $ (see Fig.~\ref{setup}) with a linear, but otherwise generic, contact condition between its faces $\Gamma^{\pm}$.  For instance the fracture may be partially closed (due to surface asperities), fluid-filled, or traction free. Here, $\Bcal_1$ is a ball of radius $R_1$ -- containing the sampling region i.e.~the search domain for  hidden fractures. The action of an \emph{incident plane wave} $\pff$ on $\Gamma$ results in the scattered field $\tilde{\bu}$ -- observed in the form of the total field 
\beq\lb{scat}
\bu(\bxi) ~=~ \pff(\bxi) \,+\, \tilde{\bu}(\bxi), \qquad \bxi\in \So,
\eeq 
over a closed measurement surface $\So=\partial \Bo$, where $\Bo$ is a ball of radius $R_2 \gg R_1$ centered at the origin. The reference i.e. ``background'' medium is assumed to be elastic, homogeneous, and isotropic with mass density $\rho$, shear modulus $\mu$, and Poisson's ratio $\nu$.  

\paragraph*{Dimensional platform} 

\textcolor{black}{For simplicity, all quantities in the sequel are rendered \emph{dimensionless} by taking~$\rho, \mu$, and~$R_1$ as the characteristic mass density, elastic modulus, and length, respectively.}

\paragraph*{Sensory data} 

In what follows, the time-dependent factor $e^{\textrm{i}\omega t}$ will be made implicit, where $\omega$ denotes the frequency of excitation. With such premise, the incident wavefield can be written as $\pff(\bxi) = \bb \, e^{-\text{i}k\bxi \cdot \bd}$ where $k=\omega/c$  signifies the wavenumber; $c$ is the relevant (compressional or shear) wave speed; $\bb\in\Omega$ is the polarization vector, and $\bd\in\Omega$ specifies the direction of propagation of the incident plane wave, noting that~$\Omega$ stands for a unit sphere.  For each incident plane wave specified via pair $(\bb, \bd)$, values of the total field $\bu(\bxi)$ are collected over $\So$. 

\paragraph*{Governing equations} 

\begin{figure}[Htp]
\vspace*{6mm} 
\center\includegraphics[width=0.68\linewidth]{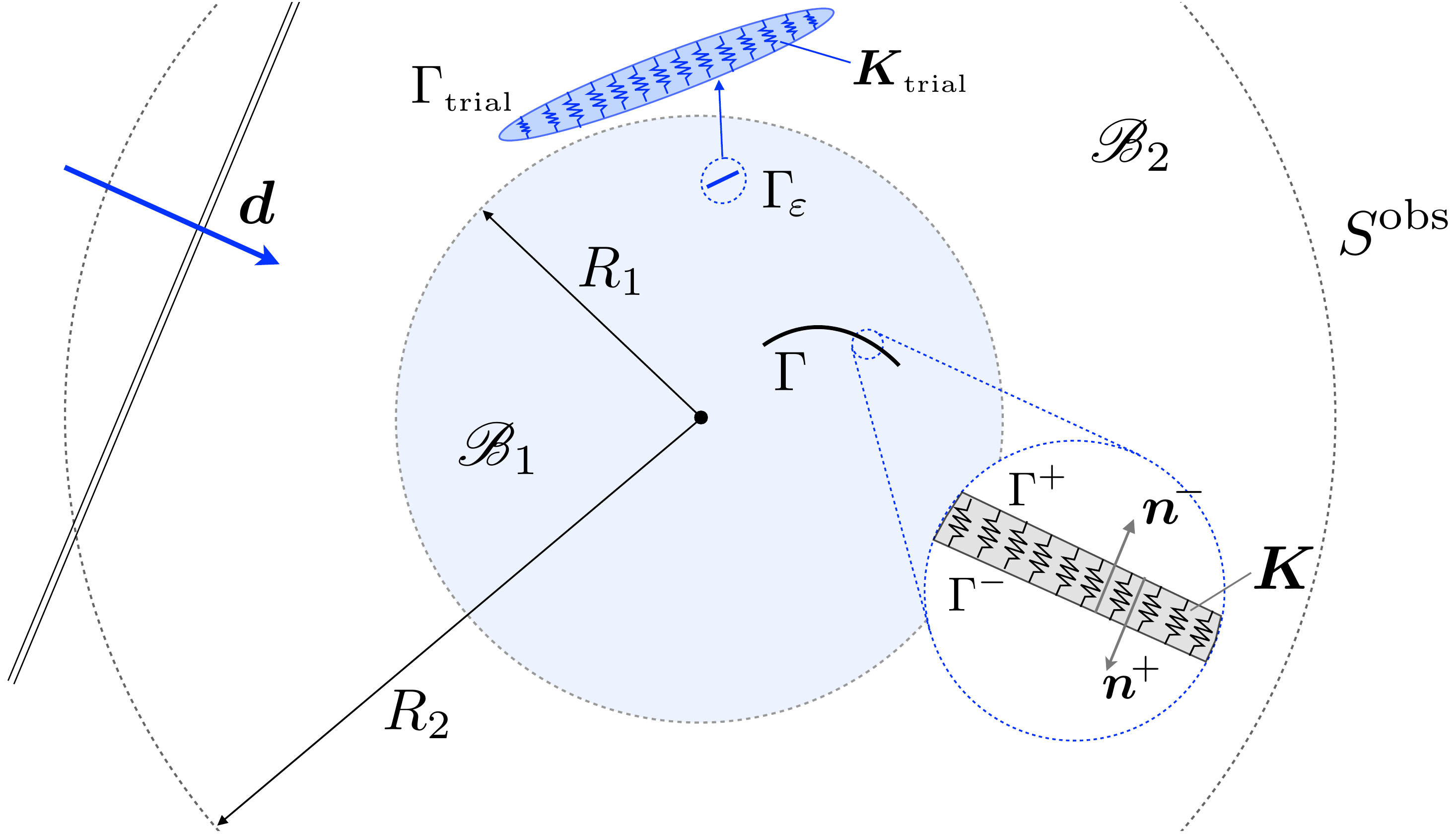} \vspace*{1mm}
\caption{Illumination of a hidden fracture $\Gamma\in\mathbb{R}^3$ with specific stiffness~$\bK$ by plane waves. Thus induced wavefield is monitored over $S\obs$.} \lb{setup}\vspace*{0mm}
\end{figure} 

With the above assumptions in place, the scattered field $\tilde{\bu}(\bxi)$ can be shown to satisfy the field equation and interfacial condition   
\beq
\begin{aligned}
\lb{ceq}  
&\nabla \sip [ \hs {\bC}\colon \! \nabla \tbu \hs ](\bxi) + \rho \exs \omega^2 \exs {\tbu}(\bxi) ~=~ \bzero, &  \quad  \bxi \in \mathbb{R}^3\backslash{\Gamma},  \\*[2mm] 
&\textcolor{black}{\tilde{\bt}^{\pm}(\bxi)} ~=~ \! \mp \exs \bK(\bxi) \exs \hs \bbu (\bxi) - \exs {\bt_f}^{\pm}(\bxi), &   \quad \bxi \in \Gamma^{\pm}, \quad \!\!
\end{aligned} 
\eeq 
complemented by the Kupradze radiation conditions~\cite{Ach2003} at infinity. Here $\llbracket {\tilde{\bu}}\rrbracket = \llbracket {\bu}\rrbracket=\bu^{+}-\bu^{-}$ signifies the crack opening displacement (COD) on~$\Gamma$; $\, \textcolor{black}{\tilde{\bt}^{\pm} = \bn^{\pm} \sip\bC\colon\!\nabla\tilde\bu^\pm}$ where $\,\bn^\pm$ is the unit normal on~$\Gamma^\pm$ (see Fig.~\ref{setup}); $\bK(\bxi)$ is a symmetric, positive-definite matrix of the specific stiffness coefficients; ${\bt_f}^{\pm} = \bn^{\pm}\!\cdot \bC\colon \! \nabla \pff$ denotes the free-field traction on~$\Gamma^\pm$, and $\bC$ is the fourth-order elasticity tensor 
\[
\bC ~=~ 2 \mu \exs \Big[\bI^{sym}_{4}+ \frac{\nu}{1-2\nu} \exs \bI_2\otimes \bI_2 \Big],
\]
in which $\bI_2$ and $\bI_{4}^{sym}\!$ stand respectively for the second-order and symmetric fourth-order identity tensors. Following the usual convention~\cite{Bonnet2005}, the \emph{unsigned} tractions and normals on a generic surface~$S$ (e.g. $\bt_f,\bn$) are referred to~$S^-$ and affiliated normal~$\bn^-$ where applicable. 

Here it is noted that $\bK$,  which accounts for the interaction between $\Gamma^{+}$ and $\Gamma^{-}$ due to e.g. surface asperities, fluid, or proppant at the fracture interface, may exhibit arbitrary spatial variations along~$\Gamma$ in terms of its normal and shear components. In light of the fact that the primary focus of this work is ``low'' frequency sensing where the illuminating wavelength exceeds most (if not all) characteristic length scales of a fracture -- de facto resulting in the spatial averaging of its properties, it is for simplicity assumed that i) the normal specific stiffness is constant along~$\Gamma$, and ii) the shear specific stiffness is both constant and isotropic~\cite{Schoenberg1980,Pyr1987,Pyrak1992}. More specifically, it is hereon assumed that 
\beq\lb{isot}
\bK(\bxi) ~=~ k_s (\be_\beta \otimes \be_\beta) + k_n (\bn \otimes \bn), \qquad \beta = 1,2, \qquad \bxi\in\Gamma,
\eeq
where $k_s=\text{const.}$ and $k_n=\text{const.}$ are the respective specific stiffnesses in the tangential $(\be_{\beta}(\bxi))$ and normal $(\bn(\bxi))$ directions of $\Gamma$; $|\be_{\beta}|=1$; $\otimes$ signifies the tensor product, and Einstein summaton convention is assumed over repeated indexes.

\paragraph*{Cost functional} 

For the purposes of solving the inverse problem the cost functional is, assuming given incident wavefield $\bu^i$, defined as 
\beq \label{Jdef}
\Jcal(\G_{\mbox{\tiny{trial}}}) ~\:=~ \int_{\So} \varphi\left(\bv,\bu\obs,\bxi\right)  \text{d}S_{\bxi},
\eeq
in terms of the least-squares misfit density 
\beq
\lb{phi_def}  \varphi(\bv,\bu\obs,\bxi) ~=~ 
\tfrac{1}{2} \: \bigl(\overline{\bv(\bksi)\!-\!\bu\obs(\bksi)}\bigr) \cdot \bW(\bxi) \cdot \bigl(\bv(\bksi)\!-\!\bu\obs(\bksi)\bigr),  
\eeq 
where $\bu\obs$ are the observations of $\bu|_{\So}$ (say polluted by noise); $\bv$ is the simulation of $\bu$ computed for trial fracture~$\G_{\mbox{\tiny{trial}}}$, and $\bW$ is a suitable (positive definite) weighting matrix, e.g.~data covariance operator.  

\section{Topological sensitivity for a fracture with specific stiffness} \label{sec3} 

\noindent On recalling~(\ref{Jdef}) and denoting 
\beq\label{geps}
\Geps ~=~ \bxio \,+\, \varepsilon\, \G_{\mbox{\tiny{trial}}}, \qquad \bxio \in \Bcal_1, 
\eeq
where $\G_{\mbox{\tiny{trial}}}$ contains the origin, the topological sensitivity (TS) of the featured cost functional can be defined as the leading-order term in the expansion of $\Jcal(\Geps)$ with respect to the vanishing (trial) fracture size, $\varepsilon\to 0$~\cite{Bellis2013}. In what follows, $\Ga$ is taken as a penny-shaped fracture of unit radius with unit normal~$\bn'$, shear specific stiffness~$\kappa_t$, and normal specific stiffness~$\kappa_n$. \textcolor{black}{To maintain the \emph{self-similarity} of~$\Gamma_{\varepsilon}$ -- in terms of mechanical behavior -- with respect to its vanishing size, (\ref{geps}) is further endowed with an $\varepsilon$-dependent matrix of specific stiffness coefficients, namely} 
\beq
\lb{keps}
\bK_\varepsilon ~=~ \varepsilon^{-1} \bK_{\mbox{\tiny{trial}}} ~=~ \frac{\kappa_s}{\varepsilon} (\be '_\beta \otimes \be '_\beta) \:+\: \frac{\kappa_n}{\varepsilon} (\bn' \otimes \bn'), 
\eeq
where $(\be'_1,\be'_2,\bn')$ constitute an orthonormal basis. Considering the mechanics of rough surfaces in contact, one may interpret~(\ref{keps}) in physical terms by noting that the height of asperities implicit to $\Ga$ is downscaled by the factor of~$\varepsilon$ when computing~$\Geps$ via~(\ref{geps}) -- see, e.g., Ch.~3 of~\cite{Sextro2007} for a phenomenological derivation of the specific contact stiffnesses from the characteristics of a rough interface. 

On the basis of the above considerations, the topological sensitivity  $\Tcal(\bxio;\bn',\kappa_n,\kappa_s)$ is obtained from the expansion 
\beq
\lb{pert}
\Jcal(\Geps) ~=~ \Jcal(\emptyset) \,+\, f(\varepsilon) \exs \Tcal(\bxio;\bn',\kappa_n,\kappa_s)\,+\, o\big(f(\varepsilon)\big) \quad \text{as} \quad \varepsilon\to 0,
\eeq
where $f(\varepsilon)\to 0$ with diminishing~$\varepsilon$, see also~\cite{Sokolowski1999,Gal2004,Guzi2004,BonnetGuzina}. Thanks to the fact that the trial scattered field $\tbv(\bxi)=\bv(\bxi)-\pff(\bxi)$ due to~$\Geps$ vanishes as $\varepsilon \rightarrow 0$, (\ref{Jdef}) can be conveniently expanded in terms of $\bv$ about $\pff$, see~\cite{Bonnet2005}. As a result (\ref{pert}) can be rewritten, to the leading order, as 
\beq
\lb{JJ0}
\Jcal(\Geps)-\Jcal(\emptyset) ~\simeq~  \int_{\So} \frac{\partial \varphi}{\partial \bv} (\pff,\bu\obs,\bxi) \cdot \tbv(\bxi) \,\, \text{d}S_{\bxi} ~=~ f(\varepsilon) \exs \Tcal(\bxio;\bn',\kappa_n,\kappa_s).
\eeq   

\paragraph*{Adjoint field approach} 

At this point, one may either differentiate~(\ref{phi_def}) at~$\bv=\bu^{i}$ and seek the asymptotic behavior of $\tbv(\bxi)$ over $ \So$, or follow the adjoint field approach~\cite[e.g.][]{Bonnet2005, Bellis2010} which transforms the domain of integration in~(\ref{JJ0}) from $\So$ to~$\Geps$ -- and leads to a compact representation of the TS formula. The essence of the latter method, adopted in this study, is to interpret the integral in~(\ref{JJ0}) through Graffi's reciprocity identity~\cite{Ach2003} between the trial scattered field $\tbv(\bxi)$ and the so-called adjoint field $\hat{\bu}(\bxi)$, whose governing equations read       
\beq\lb{2f}
\tbv \colon \!\! \left\{\bxr{ll}
\!\! \nabla \sip [ \hs { C}\colon \! \nabla \tbv \hs ](\bxi) + \rho \exs \omega^2 \exs {\tbv}(\bxi) = \bzero,  & \! \bxi \in \mathbb{R}^3\backslash{\Geps} \\*[3mm]
\!\! {\tilde{\bt}}^{\pm}(\bxi) = \! \mp \exs \bK_{\varepsilon} \exs \hs \bbv (\bxi) - \exs {\bt_f}^{\pm}(\bxi), & \! \bxi \in \Geps^{\pm} \exr \right. \!\!\!,  \quad
\hat{\bu} \colon \!\! \left\{\bxr{ll}
\!\! \nabla \sip [ \hs { C}\colon \! \nabla \hat{\bu} \hs ](\bxi) + \rho \exs \omega^2 \exs {\hat{\bu}}(\bxi) = \bzero,  & \! \bxi \in \mathbb{R}^3  \\*[3mm] 
\!\! \llbracket\hat{\bt}\rrbracket(\bxi) = \dfrac{\partial \varphi}{\partial \bv}(\pff,\bu\obs,\bxi), &   \bxi \in \So \exr \right. \!\!\!, 
\eeq  
subject to the Kupradze radiation condition at infinity. Here $\hat{\bt}$ and $\tilde{\bt}$ denote respectively the adjoint- and scattered-field tractions; $\llbracket {\tbv}\rrbracket=\tbv^{+}\!-\tbv^{-}$ is the crack opening displacement on $\Geps$, and
\[
 \llbracket\hat{\bt}\rrbracket(\bxi) ~=~ \lim_{\eta\to 0} \bn(\bxi) \sip \bC \sip \big(\nabla\hat\bu(\bxi-\eta\bn)-\nabla\hat\bu(\bxi+\eta\bn) \big), \qquad \bxi\in\So 
\]
denotes the jump in adjoint-field tractions across~$\So$ with outward normal~$\bn$. Note that the adjoint field is defined over the intact reference domain, whereby $\hat{\bu}$ is continuous $\forall\exs\bxi\in\Bcal_1$ and consequently  ${\hat{\bt}}^{\pm} \!= \mp \exs{\hat{\bt}}$ on $\Geps^{\pm}$. As a result, application of the reciprocity identity over~$\mathbb{R}^3\backslash\Geps$ can be shown to reduce~(\ref{JJ0}) to     
\beq
\lb{R-ID}
\Tcal(\bxio;\bn',\kappa_n,\kappa_s) ~=~  \big(f(\varepsilon)\big)^{-1} \exs \int_{\Geps} \hat{\bt}(\bxi) \cdot \bbv(\bxi) \,\, \text{d}S_{\bxi}. 
\eeq 
  
\subsection{Asymptotic analysis} 

\noindent Considering the trial scattered field $\tbv(\bxi)$, the leading-order contribution of the COD is sought on the boundary of the vanishing crack ($\bxi \in \Geps$) as $\varepsilon \rightarrow 0$. For problems involving kinematic discontinuities such as that investigated here, it is convenient to deploy the traction BIE framework~\cite{Bon1999} as the basis for the asymptotic analysis, namely 
\beq
\begin{aligned}
\lb{BIE}  
\bt_{\nxs f}(\bxi) - \bK_{\varepsilon} \! \cdot \nxs \bbv(\bxi) ~=~ & \bn' \sip \exs \bC : \dashint_{\Geps} \bSig(\bxi,\bx,\omega) \nxs : \nxs \boldsymbol{D}_{\!\bx} \bbv (\bx) \, \text{d}S_{\bx} \\*[1mm] 
&- \rho \exs \omega^2 \exs \bn' \sip \exs \bC \colon \!\!\! \int_{\Geps} \bU(\bxi,\bx,\omega) \sip \big( \bbv \nxs \otimes \nxs \bn' \big)(\bx) \, \text{d}S_{\bx},  \qquad    \bxi \in \Geps, 
\end{aligned} 
\eeq 
where
\[
\bU = U_{i}^{k} (\bxi,\bx,\omega) \, \be_k\otimes\be_i, \qquad 
\bSig = \Sigma_{ij}^{k}(\bxi,\bx,\omega) \, \be_k\otimes\be_i\otimes\be_j;
\]
$U^k_i(\bxi,\bx,\omega)$ and $\Sigma^{k}_{ij}(\bxi,\bx,\omega)$ (given in~\ref{App-A}) denote respectively the elastodynamic \emph{displacement} and \emph{stress} fundamental solution due to point force acting at $\bx\in\mathbb{R}^3$ in direction~$k$; $\,\dashint$~signifies the Cauchy-principal-value integral, and $\boldsymbol{D}_{\!\bx}$ is the tangential differential operator~\cite{Bon1999} on~$\Geps$, given by
\[
\boldsymbol{D}_{\!\bx}(\boldsymbol{f}) = D_{kl} (f_m)   \, \be_l\otimes\be_m\otimes\be_k, \qquad 
D_{kl}(f_m)  = n'_k f_{m,l} - n'_l f_{m,k}, 
\]
such that $n'_k=n'_k(\bx)$ and~$f_{m,k}=\partial f_m/\partial x_k$. 

\paragraph*{Scaling considerations} \lb{plt}

To facilitate the analysis of~(\ref{BIE}), all quantities are hereon assumed to be dimensionless. This is accomplished by taking the radius $R_1$ of ball $\mathcal{B}_1$ containing the sampling region (see Fig.~\ref{setup}), the mass density $\rho$, and the shear modulus~$\mu$ as the reference scales for length, mass density, and stress. 
Next, a change of variable $\bx = \bxio  +  \varepsilon \exs \bar{\bx}$ is introduced on~$\Geps$ as motivated by~(\ref{geps}), which results in the scaling relations          
\beq
\lb{scaling2}
\text{d}S_{\bx}~=~\varepsilon^2 \exs \text{d}S_{\bar\bx}, \qquad \boldsymbol{D}_{\!\bx} (\cdot)~=~ \varepsilon^{-1} \exs \boldsymbol{D}_{\!\bar\bx}(\cdot), \qquad \bt_{\nxs f}(\bxi)~=~\bt_{\nxs f}(\bxio) + O(\varepsilon), \quad \bxi \in \Geps, 
\eeq
when $\varepsilon\to 0$. In this setting, the elastodynamic fundamental tensors in~(\ref{BIE}) are known to have the asymptotic behavior 
 \beq
\lb{scaling3}
\bSig(\bxi,\bx,\omega)~=~\varepsilon^{-2} \exs \breve{\bSig}(\bar\bxi,\bar\bx)+O(1), \qquad \bU(\bxi,\bx,\omega)~=~\varepsilon^{-1} \exs \breve{\bU}(\bar\bxi,\bar\bx)+O(1), \qquad \bar\bx,\bar\bxi \in \Ga, 
\eeq
in which $\breve{\bU}(\bar\bxi,\bar\bx)$ and $\breve{\bSig}(\bar\bxi,\bar\bx)$ signify the displacement and stress tensors associated with the Kelvin's \emph{elastostatic} fundamental solution~\cite{Bon1999}. Following the logic of earlier works~\cite[e.g.][]{Bellis2013,Bellis2009,Bonnet2005}, the asymptotic behavior of a vanishing scattered field 
\beq\lb{bbb}
\bbb(\bar\bxi) ~:=~ \bbv(\bxio\!+\nxs \varepsilon \exs \bar\bxi)
\eeq 
as  $\varepsilon\to 0$ can be exposed by substituting~(\ref{scaling2})--(\ref{scaling3}) into~(\ref{BIE}) which yields
\beq \begin{aligned}
\lb{BIE2}  
\bt_{\nxs f}(\bxio) - \varepsilon^{-1} \bK_{\mbox{\tiny{trial}}} \! \cdot \nxs \bbb(\bar\bxi) ~=~ & \varepsilon^{-1} \exs \bn' \sip \exs \bC : \dashint_{\Ga}\!\! \breve\bSig(\bar\bxi,\bar\bx) \nxs : \nxs \boldsymbol{D}_{\!\bar\bx} \exs \bbb (\bar\bx) \, \text{d}S_{\bar\bx} ~+~ O(\varepsilon),  \quad~    \bar\bxi \in \Ga, 
\end{aligned} 
\eeq 
and seeking the balance of the featured leading terms~\cite{Nyfeh1993}. To solve~(\ref{BIE2}), consider a representation of the fracture opening displacement as 
\beq
\lb{COD} 
\bbb (\bar\bxi) ~\simeq~ \varepsilon^{\text{a}} \, \sigma^{f}_{ij}(\bxio) \, 
\llbracket \bV \rrbracket^{ij} (\bar\bxi), \qquad \bar\bxi \in \Ga, 
\eeq
where $\sigma_{ij}^f$ are the components of the free-field stress tensor $\bfsig_{\!f} = \bC \colon \! \nabla\bu^i$, and~$\llbracket \bV \rrbracket^{ij}$ $(i,j=1,2,3)$ are canonical solutions to be determined. On recalling that the free-field traction $\bt_{f}=\bn'\cdot\bfsig_{\!f}$ in~(\ref{BIE2}) is independent of~$\varepsilon$, one immediately finds that~$\text{a}=1$ in~(\ref{COD}) which reduces~(\ref{BIE2}) to 
\beq
\lb{canonic} 
\tfrac{1}{2} \exs \bn' \sip(\be_i \otimes \be_j +\be_j \otimes \be_i) - \bK_{\mbox{\tiny{trial}}} \! \cdot\dbbV^{ij}(\bar\bxi) ~=~  \bn' \sip \exs \bC : \dashint_{\Ga}\!\! \breve\bSig(\bar\bxi,\bar\bx) \nxs : \nxs \boldsymbol{D}_{\!\bar\bx} \exs \dbbV^{ij} \nxs (\bar\bx) \, \text{d}S_{\bar\bx} , \qquad \bar\bxi \in \Ga. 
\eeq
By analogy to the BIE formulation for an exterior (traction-free) crack problem in \emph{elastostatic}~\cite[e.g.][]{Bon1999,Bellis2010}  one recognizes that, for given pair $(i,j)$, integral equation~(\ref{canonic}) governs the fracture opening displacement $\dbbV^{ij}$ due to tractions $\pm\tfrac{1}{2} \,\exs \bn' \sip(\be_i \otimes \be_j +\be_j \otimes \be_i)$ applied to the faces $\Ga^{\pm}$ of a ``unit'' fracture~$\Ga$  with interfacial stiffness~$\bK_{\mbox{\tiny{trial}}}$ in an infinite elastic solid (recall that $\bn^{\prime\pm}=\mp\bn'$). Owing to the symmetry of $\dbbV^{ij}$ with respect to~$i$ and~$j$, (\ref{canonic}) can accordingly be affiliated with six canonical elastostatic problems in~$\mathbb{R}^3$.   

\paragraph*{The TS formula} 

Having~$\bbv$ characterized to the leading order, one finds from~(\ref{R-ID}), (\ref{scaling2}) (\ref{bbb}) and~(\ref{COD}) that $ f(\varepsilon)=\varepsilon^3$ and consequently  
\beq
\lb{TS}
\Tcal(\bxio;\bn',\kappa_n,\kappa_s) ~=~  \bfsig_{\!f}(\bxio) \colon \! \bA \colon \hat\bfsig(\bxio), \qquad 
\bA ~=~  \be_i \otimes \be_j \otimes \bigg(\int_{\Ga}  \!\dbbV^{ij}(\bar\bx) \, \text{d}S_{\bar\bx}\bigg) \otimes \bn', 
\eeq
where $\hat\bfsig=\bC:\nabla\hat{\bu}$ denotes the adjoint-field stress tensor, and~$\bA$ is the so-called \emph{polarization tensor} -- independent of~$\bxio$ and $\omega$ -- whose evaluation is examined next.   

\subsection{Elastic polarization tensor} \lb{PLT}

\noindent In prior works on the topological sensitivity~\cite[e.g.][]{Ammari2004,Ivan2007,Bellis2009,Park2012}, relevant polarization tensors were calculated analytically thanks to the available closed-form solutions for certain (2D and 3D) elastostatic exterior problems -- e.g. those for a penny-shaped crack, circular hole, and spherical inclusion in an infinite solid. To the authors' knowledge, however, analytical solution to~(\ref{canonic}) is unavailable. As a result, numerical evaluation of~$\dbbV^{ij}$ and thus~$\bA$ is pursued within a BIE framework~\cite{Pak1999,Bon1999}. Consistent with the assumed dimensional platform, the computation is effected assuming i) penny-shaped fracture of unit radius $\Lambda\!=\!1$, ii) elastic medium with unit shear modulus and mass density ($\mu=1$, $\rho=1$), and iii) various combinations between the trial fracture parameters~$(\kappa_s,\kappa_n)$ and the Poisson's ratio~$\nu$ of the background elastic solid, $\mathbb{R}^3$. In this setting, general known properties of polarization tensors are deployed along with optimization techniques to decipher the numerical results into a mathematical expression for $\bA$ which will be used later to speculate the fracture boundary condition from the TS distribution.

To solve~(\ref{canonic}) for given $\Ga \!\in \mathbb{R}^3$, A BIE computational platform is developed on the basis of the regularized \emph{traction} boundary integral equation~\cite{Bon1999} where the featured weakly-singular integrals are evaluated via the mapping techniques in~\cite{Pak1999}. Without loss of generality, it is assumed that the origin of $\bar\bxi$ coincides with the center of a penny-shaped fracture suface, and that $\bn' ={\bar\be}_3$ (see Fig.~\ref{mesh}). A detailed account of the adopted BIE framework, including the regularization and parametrization specifics, is provided in~\ref{App-B}. 

To validate the computational developments, Fig.~\ref{COD_fig}(a) compares the numerically-obtained nontrivial components of $\llbracket V_k\rrbracket^{i3}$  with their analytical counterparts along the line of symmetry in~$\Ga$ assuming traction-free interfacial conditions ($\kappa_n\nes=\nes\kappa_s\nes=\nes 0)$ and $\nu = 0.35$. Here is noted that, thanks to the problem symmetries, the variation of $\llbracket V_1\rrbracket^{13}$ along the~$\bar{\xi}_1$-axis equals that of  $\llbracket V_2\rrbracket^{23}$ along the~$\bar{\xi}_2$-axis. To illustrate the influence of~$\bK_{\mbox{\tiny{trial}}}$ on the result, Fig.~\ref{COD_fig}(b) shows the effect of shear specific stiffness $\kappa_s$ (assuming~$\kappa_n\nes=\nes 0$) on the tangential fracture opening displacement $\llbracket V_1\rrbracket^{13}$; a similar behavior is also observed concerning the effect of~$\kappa_n$ on the normal opening component $\llbracket V_3\rrbracket^{33}$.  

\paragraph*{Structure of the polarization tensor} Before proceeding further, it is useful to observe from~(\ref{TS}) that only the part of~$\bA$ with minor symmetries enters the computation of~$\Tcal$ thanks to the symmetry of~$\bfsig_{\!f}$ and~$\hat\bfsig$. Further, as shown in~\cite{Ammari2007,Bellis2013}, properties of the effective polarization tensor (hereon denoted by~$\bA^{{\text{\tiny eff}}}$) can be extended to include the major symmetry. With reference to the local basis~$(\bar{\be}_1,\bar{\be}_1,\bar{\be}_3\!=\bn')$, on the other hand, one finds that: i) $\dbbV^{\alpha\beta}\!=0$ ($\alpha,\beta\nes=\nes1,2$) due to a trivial forcing term in~(\ref{canonic}); ii) $\dbbV^{33}\propto \bar{\be}_3$ owing to the symmetry (about the $\bar\xi_3\!=0$ plane) of the boundary value problem for a penny-shaped fracture in~$\mathbb{R}^3$ solved by~(\ref{canonic}), and iii) $\dbbV^{\alpha3}\nes\cdot\bar{\be}_3\!=\dbbV^{3\alpha}\nes\cdot\bar{\be}_3\!=0$ due to the anti-symmetry of the germane boundary value problem about the $\bar\xi_3\!=0$ plane -- combined with the axial symmetry of $\Ga$ about the $\bar\xi_3$-axis. A substitution of these findings immediately verifies that $\dbbV^{3\alpha}$ and~$\dbbV^{\alpha3}$ are independent of~$\kappa_n$, whereas~$\dbbV^{33}$ does not depend on~$\kappa_s$.  Note, however, that the above arguments are predicated upon the diagonal structure of~$\bK_{\mbox{\tiny{trial}}}$ according to~(\ref{isot}). As a result, one finds that the effective polarization tensor, superseding~$\bA$ in~(\ref{TS}), permits representation 
\beq
\lb{PLTS}
\bA^{{\text{\tiny eff}}} ~=~ \alpha_s (\kappa_s,\nu) \sum_{\beta=1}^{2} (\bar\be_3\! \otimes \nxs \bar\be_\beta+\bar\be_\beta\! \otimes \nxs \bar\be_3) \otimes (\bar\be_3\! \otimes \nxs \bar\be_\beta+\bar\be_\beta\! \otimes \nxs \bar\be_3) + \alpha_n (\kappa_n,\nu) \exs (\bar\be_3\! \otimes \nxs \bar\be_3\! \otimes \nxs \bar\be_3\! \otimes \nxs \bar\be_3). 
\eeq 
To evaluate the dependency of~$\alpha_s$ and~$\alpha_n$ on their arguments, $\bA$ is evaluated numerically according to~(\ref{TS}) for various triplets~$(\nu, \kappa_s,\kappa_n)$. On deploying the Matlab optimization toolbox, the coefficients $\alpha_s (\kappa_s,\nu)$ and $\alpha_n (\kappa_n,\nu)$ are found to be rational functions of their arguments, identified as  
\beq
\lb{alpha_SN}
\alpha_s (\kappa_s,\nu)~=~\frac{4(1-\nu^2)}{3(2-\nu)(\kappa_s \textcolor{black}{\Lambda}+\nu+1)}, \qquad \alpha_n (\kappa_n,\nu)~=~\frac{8(1-\nu)(2\nu+1)}{3(\kappa_n \textcolor{black}{\Lambda}+2\nu+1)}, 
\eeq 
\textcolor{black}{where the implicit scaling parameter, $\Lambda\!=\!1$, is retained to facilitate the forthcoming application of~(\ref{alpha_SN}) to penny-shaped fractures of non-unit radius}. Assuming $\nu=0.35$, the behavior of~$\alpha_s$ and~$\alpha_n$ according to~(\ref{alpha_SN}) is plotted in Fig.~\ref{COD_fig}(c) versus the germane specific stiffness, with the corresponding numerical values included as dots. For completeness, a comparison between~(\ref{alpha_SN}) and the BIE-evaluated values of~$\alpha_s$ and~$\alpha_n$ is provided in Fig.~\ref{Pp} for a range Poisson's ratios, $\nu\in[0.05,0.45]$.    
\begin{figure}[!h]
\vspace*{0.0mm} 
\center\includegraphics[width=1\linewidth]{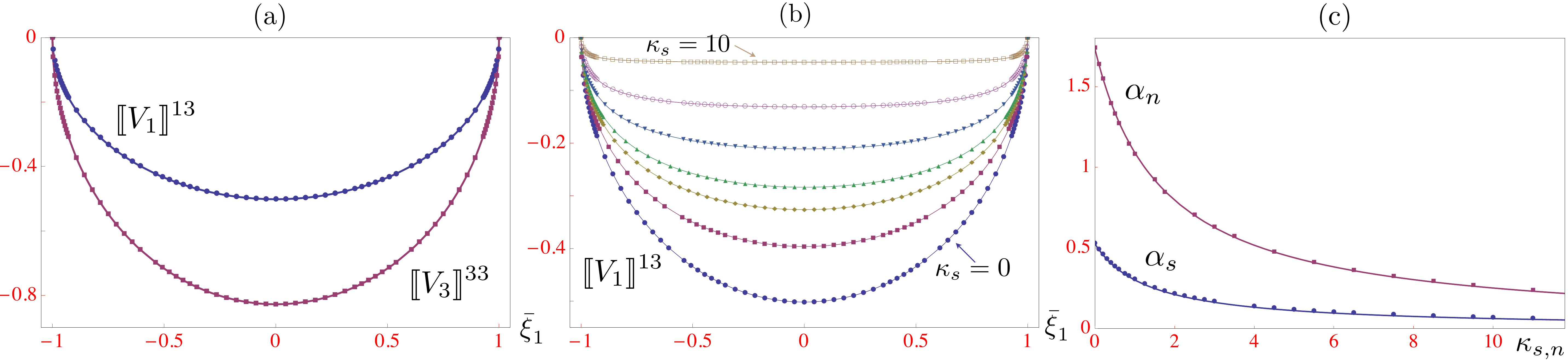} \vspace*{-4mm} 
\caption{Dependence of the crack opening displacement~$\dbbV^{ij}$ and coefficients of the effective polarization tensor~$\bA^{{\text{\tiny eff}}}$ on the specific stiffnesses, $\kappa_s$ and~$\kappa_n$, of a trial fracture assuming $\nu=0.35$: (a) analytical values (solid lines) \emph{vs.}~numerical values (dots) of the COD along $(\bar\xi_1,0,0)$ for a traction-free crack; (b) evolution of the shear COD  $\llbracket V_1\rrbracket^{13}$ with increasing $k_s$, and (c) proposed variation~(solid lines)~\emph{vs.}~numerical variation (dots) of $\alpha_{n}$ and~$\alpha_{n}$~versus the relevant specific stiffness. } \lb{COD_fig}\vspace*{0mm} 
\end{figure} 

From~(\ref{PLTS}), it is seen that the interfacial condition on~$\Ga$ has no major effect on the \emph{structure} of the polarization tensor. This may explain the observation from numerical experiments (Section~\ref{sec5}) that by using a traction-free trial crack ($\kappa_s\nes=\kappa_n\nes=0$) in~(\ref{PLTS}), the \emph{geometrical} characteristics of a hidden fracture -- its location, normal vector and, in the case of high excitation frequencies, its shape -- can be reconstructed regardless of the assumed interfacial condition. Nonetheless, a trial crack with interacting surfaces introduces two new parameters ($\kappa_s$ and $\kappa_n$) to the reconstruction scheme, whereby further information on the hidden fracture's interfacial condition may be extracted. In this vein, it can be shown that the first-order topological sensitivity~(\ref{TS}) is, for given~$\bxio$, a monotonic function of $\kappa_{n}, \, \kappa_{s}$; accordingly, precise contact conditions on $\Gamma$ (the true fracture) cannot be identified separately via e.g.~a TS-based minimization procedure. In principle, such information could be retrieved by pursuing a higher-order TS scheme~\cite[e.g.][]{Bonnet2011} which is beyond the scope of this study. As shown in the sequel, however, the present (first-order) TS sensing framework is capable of qualitatively \emph{identifying the ratio} between the specific shear and normal stiffnesses on~$\Gamma$ -- an item that is of great interest in e.g. hydraulic fracturing applications~\cite[e.g][]{Bakulin2000,Verdon2013}. 

\begin{figure}[!h]
\vspace*{0mm} 
\center\includegraphics[width=1\linewidth]{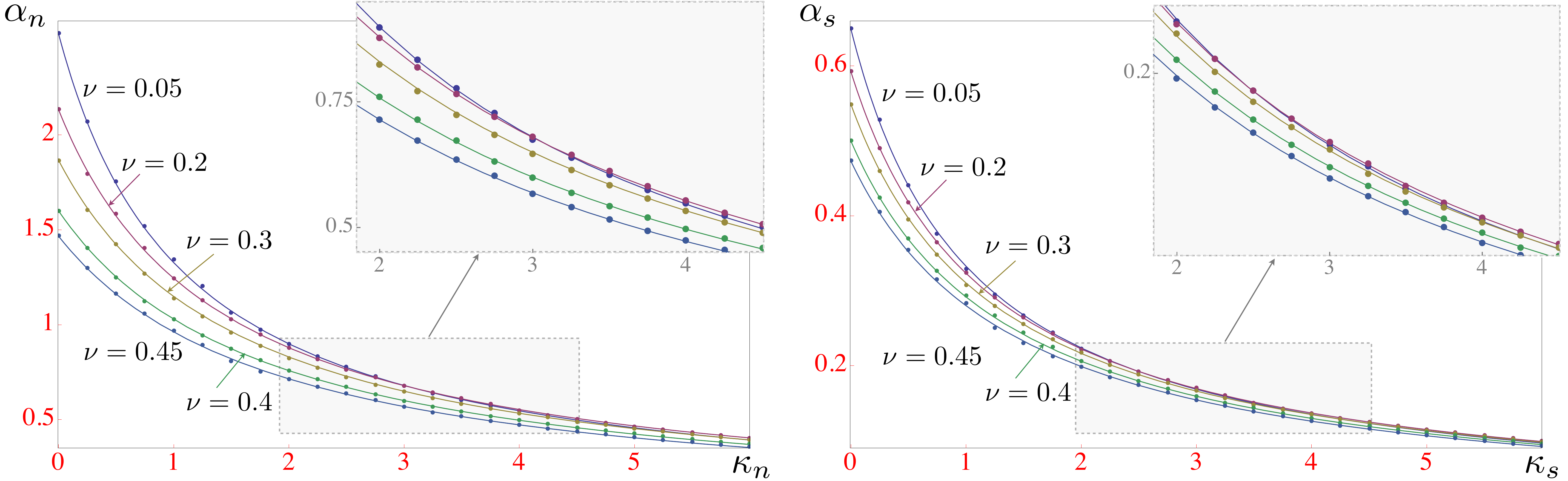} \vspace*{-4mm} 
\caption{Performance of the formulas~(\ref{alpha_SN}) describing $\alpha_n(\kappa_n,\nu)$ and $\alpha_s(\kappa_s,\nu)$ in the $(\nu, \kappa_s,\kappa_n)$-space: numerical values (dots) \emph{vs.} proposed expressions (solid lines).} \lb{Pp}\vspace*{0.0mm} 
\end{figure} 

\section{Qualitative identification of the fracture's interfacial condition} \label{sec4} 

\noindent In this section, an ability of the TS indicator function~(\ref{TS}) to qualitatively characterize the interfacial condition of \emph{nearly-planar fractures} is investigated in the ``low'' frequency regime, where the illuminating wavelength exceeds the length scales spanned by~$\Gamma$. Such limitations are introduced to facilitate the asymptotic analysis where the true (finite) fracture is approximated as being penny-shaped. The proposed analysis naturally extends to arbitrary-shaped, near-flat fractures by assuming a diagonal, \emph{to the leading order}, interfacial stiffness matrix and the COD profile proposed in~\cite{Fabrikant1987} to calculate the relevant polarization tensor required in the asymptotic approximation of the scattered field due to~$\Gamma$. In the approach, it is also assumed that the fracture location and normal vector to its surface are identified via an initial TS reconstruction performed using a traction-free trial crack ($\kappa_s\nes=\kappa_n\nes=0$) as described in Section~\ref{sec5}. As shown there, however, the latter geometric reconstruction is not limited to a particular frequency range.  
 
On recalling the least-squares distance function~$\varphi$ in~(\ref{phi_def}) and the leading-order perturbation of $J(\emptyset)$ in~(\ref{JJ0}), the TS may be rewritten as
\beq
\lb{R_TS}
\varepsilon^3 \, \Tcal(\bxio;\bn',\kappa_n,\kappa_s) \,= - \text{Re} \Big[ \nxs \int_{\So} \!\! \tbu^*(\bxi) \sip \exs \tbv(\bxi) \, \text{d}S_{\bxi} \Big],
\eeq
where~$(\cdot)^*$ signifies complex conjugation and $-\tbu^*(\bxi)=\partial\varphi(\bv,\bu\obs,\bxi)/\partial\bv|_{\bv = \bu^{i}}$, noting that $\tbu=\bu-\bu^i$ is the scattered field due to~$\Gamma$, see~(\ref{scat}). Given the fact that the hidden fracture is separated from the observation surface, the scattered field on~$\So$ can be expressed via a displacement boundary integral representation as 
\beq
\lb{True_SC}
\tbu(\bxi)=\int_{\Gamma} \big(\bbu(\bx)\otimes\bn\big) \colon \nxs \bSig(\bxi,\bx,\omega) \,\, \text{d}S_{\bx}, \qquad \bxi \in \So.
\eeq

\paragraph*{Small crack asymptotics} 

Consider the testing configuration as in Fig.~\ref{setup}, and let $\Gamma$ with interfacial stiffness~(\ref{isot}) be illuminated by a ``low-frequency'' plane wave $\bu^i\!=\nes\bb \, e^{-\text{i}k\bxi \cdot \bd}$ in that $k\nes L\ll 1$, where~$2L$ is the characteristic size of~$\Gamma$. With such premise and hypothesis that $2L\nes<\nes R_1\nes\ll\nes R_2$ made earlier (recall that~$R_2$ is the radius of~$\So$), the hidden fracture can -- in situations of predominantly flat geometry and an~$O(1)$ aspect ratio -- be approximated as a penny-shaped fracture of finite radius $L$.  In this setting (\ref{True_SC}) can be expanded, utilizing the developments from Section~\ref{sec3}, as 
\beq
\lb{True_SC_app}
\tbu(\bxi) ~\simeq \bigg(\int_{\Gamma} \big(\bbu(\bx)\otimes\bn\big) \, \text{d}S_{\bx}\bigg) \colon \nxs \bSig(\bxi,\bz,\omega) ~\simeq~
L^3 \, \bfsig^f(\bz)  \colon \nxs \bA^{\!\Gamma} \colon \! \bSig(\bxi,\bz,\omega), \qquad \bxi \in \So,
\eeq
where $\bz$ (to be determined) is an indicator of the fracture location, and~$\bA^{\!\Gamma}$ is its effective (low-frequency) polarization tensor given by 
\beq
\lb{PLTS2}
\bA^{\!\Gamma} \,=~ \sum_{\beta=1}^{2} \varsigma_s \exs (\bn \! \otimes \nxs \be_\beta+\be_\beta\! \otimes \nxs \bn) \otimes (\bn\! \otimes \nxs \be_\beta+\be_\beta\! \otimes \nxs \bn) \,+\, \varsigma_n \exs (\bn \! \otimes \nxs \bn \! \otimes \nxs \bn\! \otimes \nxs \bn).
\eeq 
Here~$\bn$ is the normal on~$\Gamma$; $(\be_1,\be_2,\bn)$ make an orthonormal basis, and 
\beq
\lb{varsigma_SN}
\varsigma_s (k_s,\nu)~=~\frac{4(1-\nu^2)}{3(2-\nu)(k_s L+\nu+1)}, \qquad \varsigma_n (k_n,\nu)~=~\frac{8(1-\nu)(2\nu+1)}{3(k_n L+2\nu+1)}, 
\eeq 
where~$k_s$ and~$k_n$ are the shear and normal specific stiffness of the true fracture according to~(\ref{isot}), see also (\ref{keps}) and~(\ref{alpha_SN}). On taking without loss of generality $(\be_1,\be_2,\bn)$ as the basis of the global coordinate system, (\ref{True_SC_app}) can be rewritten in component form as 
\beq
\lb{TSC}
\tilde{u}_j(\bxi) ~=~ L^3 \, \big[4 \exs \varsigma_s(k_s,\nu)  \exs \Sigma^j_{3\beta}(\bxi,\bz,\omega) \, \sigma^f_{3\beta}(\bz) +  \varsigma_n(k_n,\nu)  \exs \Sigma^j_{33}(\bxi,\bz,\omega) \, \sigma^f_{33} (\bz) \big], \qquad \bxi \in \So,
\eeq
where $\beta\nes=\nes 1,2$ and the summation is assumed over repeated indexes as before. 

Assuming that the location and (average) normal on~$\Gamma$ are identified beforehand, one may set~$\bn'=\bn$ and $\bxio=\bz$ in~(\ref{R_TS}) and expand the scattered field due to vanishing \emph{trial} fracture~$\Geps=\bxio+\varepsilon \G_{\mbox{\tiny{trial}}}$ in an analogous fashion as 
\beq
\lb{Trial_SC}
\tilde{v}_j(\bxi) ~=~ \varepsilon^3 \, \big[4 \exs \alpha_s(\kappa_s,\nu)  \exs \Sigma^j_{3\beta}(\bxi,\bz,\omega) \, \sigma^f_{3\beta}(\bz) +  \alpha_n(\kappa_n,\nu)  \exs \Sigma^j_{33}(\bxi,\bz,\omega) \, \sigma^f_{33} (\bz) \big], \qquad \bxi \in \So,
\eeq
On substituting~(\ref{TSC}) and~(\ref{Trial_SC}) into~(\ref{R_TS}), the leading-order TS contribution in the low-frequency regime is obtained as 
\beq
\lb{TS_str}
\Tcal(\bz,\kappa_n,\kappa_s) ~\simeq\: - L^3 \exs \big[\alpha_n \exs \varsigma_n  Q_1(\bz) \,+\, \big(\alpha_s\exs\varsigma_n + \alpha_n\exs\varsigma_s\big) Q_2(\bz) \,+\, \alpha_s \exs \varsigma_s  Q_3(\bz) \big],
\eeq
where
\beq
\lb{mathC}
\begin{aligned}
&Q_1(\bz) ~=~ |\sigma_{33}^f(\bz)|^2 \nxs \int_{\So} \!\!  [\Sigma^{j*}_{33} \exs \Sigma^j_{33}](\bxi,\bz,\omega) \, \text{d}S_{\bxi}, \\*[1mm]
&Q_2(\bz) ~=~ 4 \exs \text{Re} \Big\{ [\sigma_{33}^{f*} \exs \sigma_{3\beta}^f](\bz) \nxs \int_{\So} \!\!  [\Sigma^{j*}_{33}  \exs \Sigma^j_{3\beta}](\bxi,\bz,\omega) \, \text{d}S_{\bxi} \Big\}, \\*[1mm]
&Q_3(\bz) ~=~ 16 \exs \text{Re} \Big\{ [\sigma_{3\alpha}^{f*} \exs \sigma_{3\beta}^f](\bz) \nxs \int_{\So} \!\!  [\Sigma^{j*}_{3\alpha}  \exs \Sigma^j_{3\beta}](\bxi,\bz,\omega) \, \text{d}S_{\bxi} \Big\}, \\*[1mm]
\end{aligned}
\eeq
where~$(\cdot)^*$ indicates complex conjugation, $\alpha,\beta=1,2$, and $j=1,2,3$ as before. With reference to~\mbox{\ref{App-A}}, integration of the anti-linear forms, featuring components of the fundamental stress tensor, over $\So$ can be performed analytically by approximating the distance $r = |\bxi-\bz|, \,\,\, \bxi \in \So$ in~(\ref{HFS}) and~(\ref{Cons}) to the leading order as~$r = R_2$. As a result, the behavior of~(\ref{mathC}) can be approximated as    
\begin{multline}
\lb{mathC2}
Q_1  \simeq  \frac{|\sigma_{33}^f(\bz)|^2}{4 \pi R_2^2}  \Big[ \tfrac{1}{5} \exs \text{X}_2+\tfrac{4}{3}\exs \text{Re}(\text{X}_3)+ |P(R_2)|^2  \Big], \quad Q_2 = O\Big(\frac{1}{R_2^3}\Big) \, \simeq \, 0, \\
\quad Q_3  \simeq  \frac{4}{ \pi R_2^2}  [\sigma_{3\beta}^{f*} \exs \sigma_{3\beta}^f ](\bz) \exs \Big[ \tfrac{1}{15} \exs \text{X}_2+\tfrac{2}{3}\exs |F(R_2)|^2 \Big],   
\end{multline} 
owing to the initial premise of ``far field'' sensing (namely $R_2\gg 1$ within the adopted dimensional platform), where $\text{F}(\cdot)$, $\text{P}(\cdot)$, $\text{X}_{2}$ and $\text{X}_{3}$ are given by~(\ref{Cons}) and~(\ref{X}) in~\mbox{\ref{App-A}} along with the details of the calculation procedure. 

A remarkable outcome of the analysis is that the coefficient $Q_2$ describing the mixed term in~(\ref{mathC2}) vanishes to the leading order, whereby the TS structure \emph{decouples} and may be perceived as a superposition of the normal and shear contributions. Specifically, (\ref{TS_str}) becomes 
\beq
\lb{TS_str2}
\Tcal(\bz;\bn,\kappa_n,\kappa_s) ~\simeq~ -L^3 \exs \big( \alpha_n(\kappa_n,\nu) \exs \varsigma_n  Q_1(\bz) \,+\, \alpha_s(\kappa_s,\nu) \exs \varsigma_s  Q_3(\bz) \big), 
\eeq
where $\varsigma_n$ and~$\varsigma_s$ are given by~(\ref{varsigma_SN}). On account of~(\ref{TS_str2}) and the limiting behavior of $\alpha_n$ (resp.~$\alpha_s$) as a function of the trial interface parameter $\kappa_{n}$ (reps.~$\kappa_{s}$) in~(\ref{alpha_SN}), the ratio between the shear and normal specific stiffness ($k_s/k_n$) of a hidden fracture can be qualitatively identified via the following procedure.    

\paragraph*{Interface characterization scheme} 

With reference to the last paragraph of Section~\ref{sec3}, it is hereon assumed that the fracture location is identified beforehand from the low-frequency scattered field data as
\beq\label{zdef}
\bz ~=~ \text{arg min}_{\bxio} \Tcal(\bxio;\bn'(\bxio),0,0)
\eeq
where, for each sampling point, $\bn'(\bxio)$ is the optimal unit normal (minimizing~$\Tcal$ at that point) as described in~\cite{Bellis2013}.  In this setting the \emph{effective} normal to a hidden fracture, $\bn$, is obtained by averaging~$\bn'(\bxio)$ over a suitable neighborhood of~$\bz$. Here it is for completeness noted that the spatial variation of~$\bn'(\bxio)$ provides a clue whether the hidden fracture is nearly-planar and thus amenable to the proposed treatment, see the low-frequency results in Figs.~\ref{LFN} and~\ref{Lcurve} as an example. 

In the vicinity of~$\bz$, the TS characterization of the fracture's interfacial condition is performed using two (vanishing) trial fractures: one allowing for the COD in the normal direction only by assuming~$(\kappa_s,\kappa_n)=(\infty,0)$, and the other restricting the COD to tangential directions via $(\kappa_s,\kappa_n)=(0,\infty)$. The resulting TS fields are then normalized by the relevant components of~$\bfsig_{\!f}(\bz)$ according to~(\ref{mathC}). Thanks to the limits $\lim_{\kappa_s\to\infty}\alpha_s=0$ and~$\lim_{\kappa_n\to\infty}\alpha_n=0$, this leads to the respective indicator functionals 
\beq
\lb{Rec_sch}
\begin{aligned}
& \Tcal_1(\bxio) ~=~ \frac{1}{|\sigma_{33}^f(\bz)|^2} \exs  \Tcal(\bxio\!;\bn,0,\infty) ~\simeq~ -  \frac{L^3 \exs \varsigma_n}{4 \pi R_2^2} \exs  \alpha_n(0,\nu)\Big[ \tfrac{1}{5} \exs \text{X}_2+\tfrac{4}{3}\exs \text{Re}(\text{X}_3)+ |P(R_2)|^2  \Big], \\*[1mm]
&  \Tcal_2(\bxio) ~=~ \frac{1\!}{[\sigma_{3\beta}^{f*} \exs \sigma_{3\beta}^f] (\bz)} \exs  \Tcal(\bxio\!;\bn,\infty,0) ~\simeq~  -\frac{4 \exs L^3 \exs \varsigma_s}{ \pi R_2^2}  \exs \alpha_s(0,\nu)\Big[ \tfrac{1}{15} \exs \text{X}_2+\tfrac{2}{3}\exs |F(R_2)|^2 \Big],
\end{aligned}
\eeq
were~$\bxio$ is in a neighborhood of~$\bz$. In practical terms, the latter is identified as a ball (centered at~$\bz$) whose radius is a fraction of the germane (compressional or shear) wavelength.  On the basis of~(\ref{Rec_sch}), one can identify three distinct interface scenarios:

\begin{description}
\item{1.} The situation where $k_s$ and $k_n$ are of the same order of magnitude (e.g.~traction-free crack), in which case $[\Tcal_1/\Tcal_2](\bxio)=O(1)$ regardless of~$\nu$. This is verified in Fig.~\ref{T1T2}, where the ratio $\Tcal_1/\Tcal_2$ is plotted against $R_2$ -- scaled by the shear wavelength $\lambda_s\nes=\nes 2\pi c_s/\omega$ -- for various Poisson's ratios and two ``extreme'' sets of interfacial stiffnesses, namely $k_\star\nes=\nes 0.1$ and $k_\star\nes=\nes 10$ $\,(\star=s,n)$. In the context of the proposed characterization scheme, Fig.~\ref{ID}(a) plots the spatial distribution of~$\Tcal_1$ and~$\Tcal_2$ in a neighborhood of a traction-free fracture. As can be seen from the display, the fracture is visible from both panels, which suggests that $k_s$ and $k_n$ are comparable in magnitude. 

\item{2.} The limiting case $k_s\!\ll\!k_n$ that, in the context of energy applications, corresponds to a hydraulically-isolated fracture~\cite{pyrak2014, Place2014, Bakulin2000}. Under such circumstances one has $\Tcal_1(\bxio)\!\ll\!\Tcal_2(\bxio)$, which can be verified by letting $\varsigma_n\to0$ in~(\ref{varsigma_SN}). This behavior is shown in Fig.~\ref{ID}(b), where a hidden fracture with~$(k_s,k_n)=(2,100)$ is visible in the distribution of~$\Tcal_2$, but not in that of~$\Tcal_1$. Note that the image of a fracture is notably smeared due to the use of low-frequency excitation, as postulated by the interface characterization scheme. 

\item{3.} The case when the fracture surface is under small normal pressure and the effect of surface roughness is significant, namely~$k_s\!\gg\!k_n$. Here  $\Tcal_1(\bxio)\!\gg\! \Tcal_2(\bxio)$ due to the fact that $\lim_{k_s\to\infty}\varsigma_s=0$ thanks to~(\ref{varsigma_SN}). This is illustrated in Fig.~\ref{ID}(c), where the fracture with~$(k_s,k_n)=(100,2)$ appears in the distribution of~$\Tcal_1$ only, suggesting that its interfacial stiffness according to~(\ref{isot}) is dominated by the shear component~$k_s$. 
\end{description}

Here it is worth noting that the above TS scheme for \emph{qualitative identification} of the ratio $k_s/k_n$ is non-iterative, and shines light on an important contact parameter at virtually no computational surcharge -- beyond the effort needed to image the fracture. In particular, since the (low-frequency) free and adjoint fields are precomputed toward initial estimation of the fracture location $\bz$ and effective normal vector $\bn$, they can re-used to compute~$\Tcal_1$ and~$\Tcal_2$ via~(\ref{TS}), (\ref{PLTS}), (\ref{alpha_SN}) and~(\ref{Rec_sch}), wherein the only variable is the effective polarization tensor~$\bA^{{\text{\tiny eff}}}$ -- describing trial fractures with different interfacial condition. 
\begin{figure}[h]
\vspace*{0mm} 
\center\includegraphics[width=1.0\linewidth]{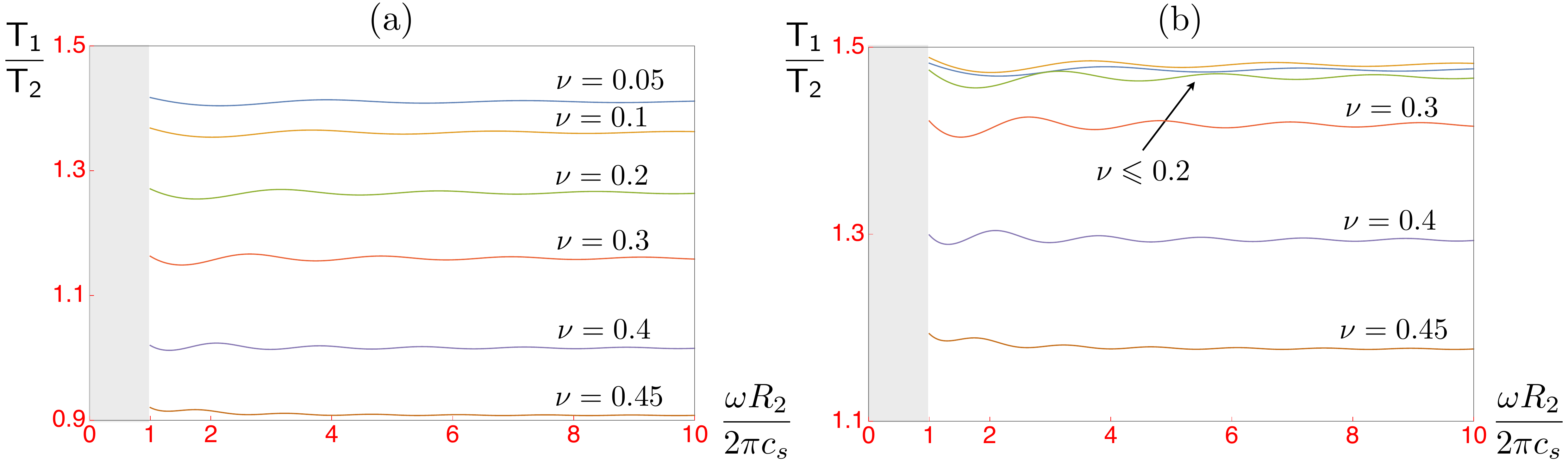} \vspace*{-5.0 mm}
\caption{Ratio $[\Tcal_1/\Tcal_2](\bz)$ \emph{vs.} the radius of $\So$, measured in shear wavelengths, assuming the fracture specific stiffnesses as (a) $k_s=k_n = 0.1$, and (b) $k_s=k_n = 10$.} \lb{T1T2}\vspace*{-2.0mm}
\end{figure}

\paragraph*{Illumination by multiple incident waves} In many situations the imaging ability of a TS indicator functional can be improved by deploying multiple illuminating wavefields, which in the context of this study translates into multiple directions~$\bd$ of plane-wave incidence. In this case the ``fortified'' TS functional can be written as
\[
\breve{\Tcal} ~=~ \int_{\Omega_{\bd}}\!\! \Tcal|_{\bd} \cdot \text{w}(\bd)\,  \exs \text{d}S_{\bd},
\]
which superimposes (in a weighted fashion) the TS distributions for incident plane waves spanning a given subset, $\Omega_{\bd}$, of the unit sphere. In the context of~(\ref{TS_str}) and (\ref{mathC}), the only $\bd$-dependent items are the components of the free-field stress tensor~$\bfsig_{\!f}$. As a result, the criteria deduced from~(\ref{Rec_sch}) remain valid under the premise of multiple incident-wave illumination provided that the free-field terms
\[
[\sigma_{33}^{f*}\exs\sigma_{33}^{f}], \quad~ [\sigma_{33}^{f*} \exs \sigma_{3\beta}^f], \quad~ 
[\sigma_{3\alpha}^{f*} \exs \sigma_{3\beta}^f]
\]
in~(\ref{mathC})
are replaced respectively by 
\[
\int_{\Omega_{\bd}} \nxs[\sigma_{33}^{f*}\exs\sigma_{33}^{f}]_{\bd} \, \text{d}S_{\bd}, \quad~ 
\int_{\Omega_{\bd}} \nxs[\sigma_{33}^{f*} \exs \sigma_{3\beta}^f]_{\bd} \, \text{d}S_{\bd}, \quad~
\int_{\Omega_{\bd}} \nxs[\sigma_{3\alpha}^{f*} \exs \sigma_{3\beta}^f]_{\bd} \, \text{d}S_{\bd}.
\]     
     
\section{Numerical results} \label{sec5} 

\noindent A set of numerical experiments is devised to illustrate the performance of the TS for elastic-wave imaging of fractures and qualitative characterization of their interfacial condition. To this end the BIE computational platform, described in~\ref{App-B}, is used to generate the synthetic data ($\bu\obs$) for the inverse problem. While the main focus of the study is on the low-frequency sensing as mandated by the characterization scheme, the results also include the TS reconstruction examples at intermediate-to-high frequencies, motivating future research in this area. The sensing setup, reflecting the assumptions made in Section~2, is shown in Fig.~{\ref{Num_setup}} where the ``true'' fracture $\Gamma$ is either i) a penny-shaped crack with diameter $L=0.7$ and normal $\bn = (0,-1/\sqrt{2},1/\sqrt{2})$ -- see Fig.~{\ref{Num_setup}}(a), or ii) a cylindrical crack of length $L=0.7$ and radius $R=0.35$ shown in Fig.~{\ref{Num_setup}}(b). The shear modulus, mass density, and Poisson's ratio of the background medium are taken as $\mu\nes=\nes1$, $\rho\nes=\nes1$ and $\nu\nes=\nes0.35$, whereby the shear and compressional wave speeds read $c_s=1$ and $c_p=2.08$, respectively. The elastodynamic field produced by the action of illuminating (P- or S-) plane wave, propagating in direction $\bd$, on $\Gamma$ is measured over~$S\obs$ -- taken as the surface of a \emph{cube} with side $3.5$ centered at the origin. This was done to investigate the robustness of the interface characterization scheme developed in Section~\ref{sec4} with respect to the simplifying assumptions used to derive~(\ref{TS_str})--(\ref{mathC2}), namely the premise that $\So$ is a \emph{sphere}. On the adopted observation surface, the density of sensing points is chosen to ensure at least four sensors per shear wavelength. In what follows, the TS is computed inside a sampling cube of side 2 centered at the origin; its spatial distribution is plotted either in three dimensions, or in the mid-section $\Pi_1$ (resp.~$\Pi_2$) of the penny-shaped  (resp.~cylindrical) fracture shown in Fig.~\ref{Num_setup}. In the spirit of an effort to test the robustness of the adopted simplifying assumptions, it is noted that the ratio between the radii of spheres circumscribing the sensing area and the sampling region is $R_2/R_1=1.75\centernot\gg 1$. 
\begin{figure}[h]
\vspace*{5mm} 
\center\includegraphics[width=1.0\linewidth]{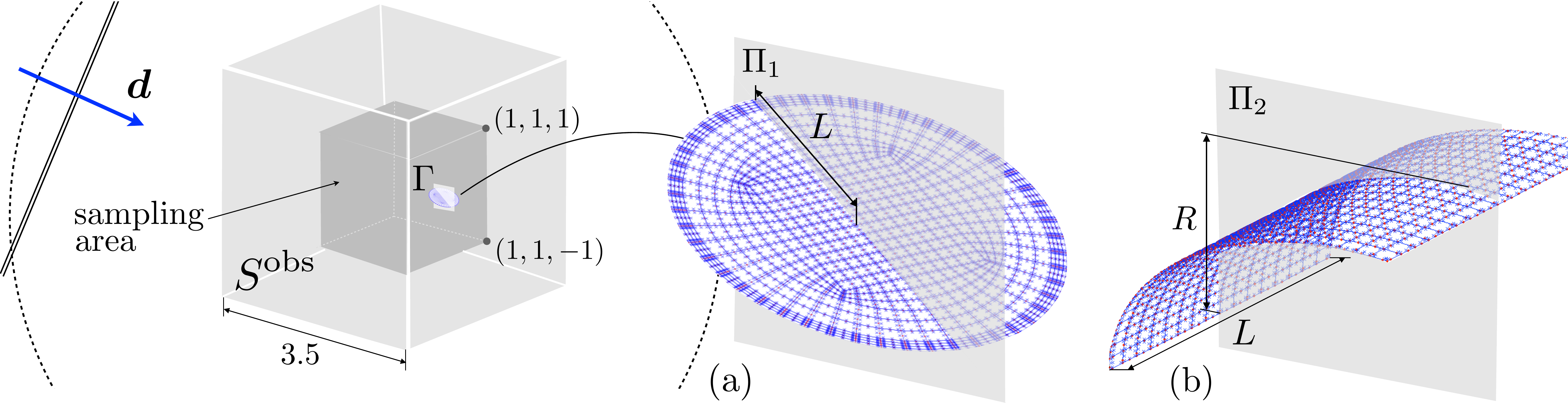} 
\caption{Model problem: (a) sensing configuration with an embedded penny shaped fracture, and (b) with non-planar scatterer i.e.~cylindrical fracture.} \lb{Num_setup}\vspace*{0.0mm}
\end{figure}

\paragraph*{Low-frequency TS reconstruction} 

Consider first the case of an ``isolated fluid-filled'' planar fracture shown in Fig.~\ref{Num_setup}(a), whose specific stiffnesses are given by $(k_s,k_n)\nxs =\nxs(2,100)$. To geometrically identify the fracture, the region of interest is illuminated by twelve P- and S- incident waves propagating in directions $\bd \in \lbrace (\pm1,0,0),(0,\pm1,0),(0,0,\pm1) \rbrace$, while assuming~$(\kappa_s,\kappa_n)\nes=\nes(0,0)$ for the (vanishing) trial fracture. The illuminating frequency is taken as $\omega\nes=\nes 4$, whereby the ratio between the probing shear wavelength \mbox{$\lambda_s\nes= 2\pi c_s/\omega\simeq 1.6$} and fracture diameter, $L$, is approximately 2.2. For each incident wave and each sampling point, the optimal normal vector~$\bn'(\bxio)$ is estimated by evaluating \mbox{$\Tcal(\bxio; \bn' =[\cos(\theta_i) \cos(\phi_j), \sin(\theta_i) \cos(\phi_j), \sin(\phi_j)],0,0)$} for trial pairs $(\theta_i,\phi_j)\in[0,2\pi]\times[0,\pi]$ and choosing the pair that minimizes $\Tcal(\bxio;\cdot)$. Thus obtained TS distributions are superimposed as
\[
\breve{\Tcal}(\bxio;\bn',0,0) ~=~ |\text{min}_{\bxio}\breve{\Tcal}|^{-1} \, \sum_{\text{n}=1}^{12} \, \Tcal(\bxio;\bn',0,0)|_{(\bb_{\text{n}}, \bd_{\text{n}})},
\] 
resulting in a composite indicator function whose spatial distribution shown in Fig.~\ref{LFN}(a).  In this setting, the fracture is geometrically identified via a region where $\breve{\Tcal}$ attains its most pronounced negative values, see Fig.~\ref{LFN}(b). To provide a more complete insight into the performance of the approach, Fig.~\ref{LFN}(c) plots the distribution of point-optimal unit normal~$\bn'(\bxio)$ over the reconstructed region. For the purposes of interface characterization, the fracture location~$\bz$ is identified according to~(\ref{zdef}), while its effective normal~$\bn'$ is computed by averaging~$\bn'(\bxio)$ over the reconstructed volume in Fig.~\ref{LFN}(b). 
\begin{figure}[h]
\vspace*{2mm} 
\center\includegraphics[width=0.9\linewidth]{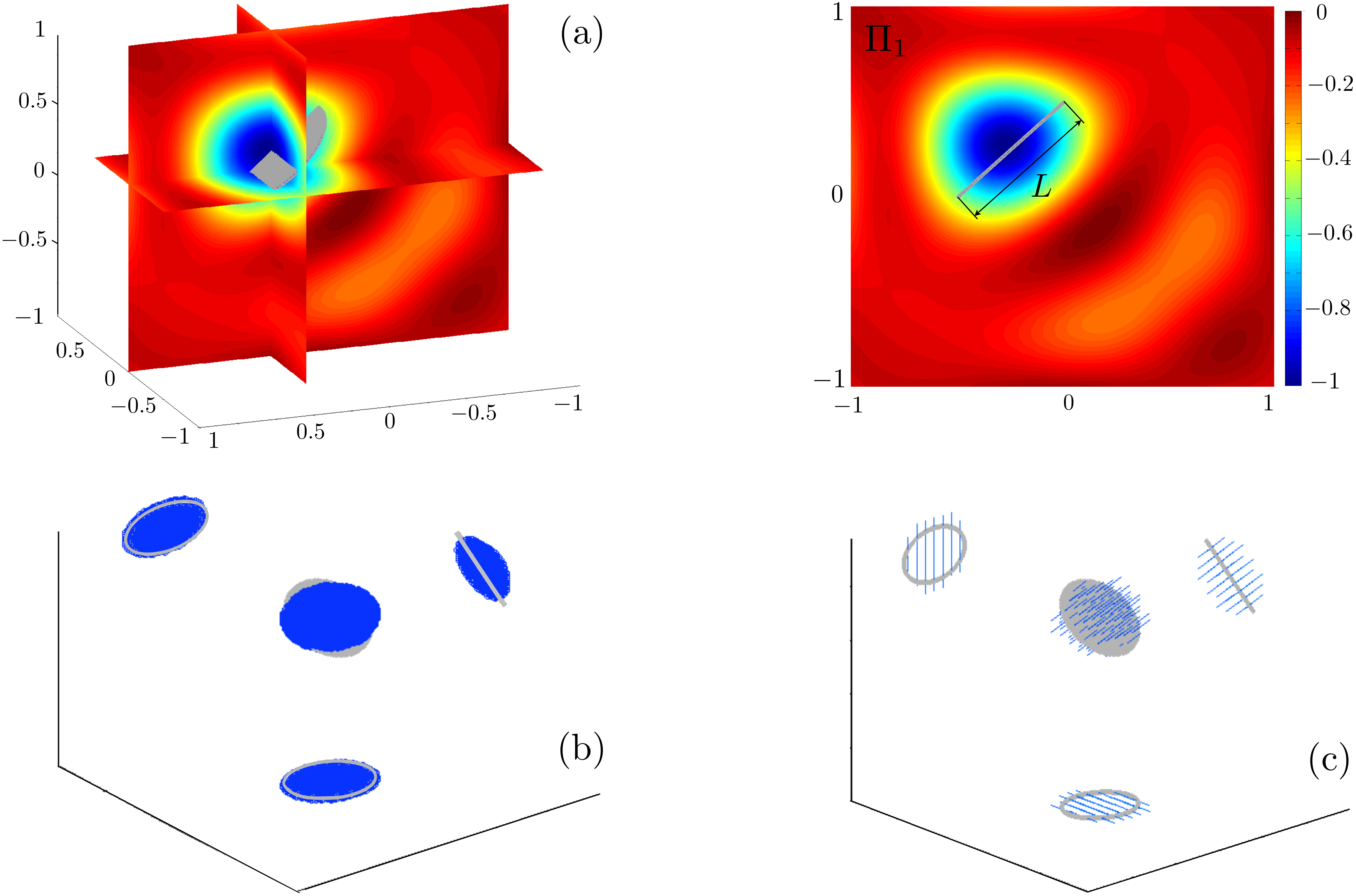} 
\caption{Low-frequency reconstruction of an ``isolated fluid-filled'' planar fracture with $(k_s,k_n)\nxs =\nxs(2,100)$: (a)~composite TS distribution $\breve{\Tcal}(\bxio;\bn',0,0)$ in three dimensions, and in the fracture mid-section $\Pi_1$; (b)~region containing the most pronounced negative TS values: $-1\nes\leqslant\nes\breve{\Tcal}\nes\leqslant\nes -0.6$, and (c)~point-optimal normal vector~$\bn'(\bxio)$ plotted over the true fracture surface $\Gamma$.} \lb{LFN}\vspace*{0.0mm}
\end{figure}

For completeness, the above-described geometrical identification procedure is next applied to the cylindrical fracture in Fig.~\ref{Num_setup}(b), assuming the ``true'' specific stiffnesses as $(k_s,k_n)\nxs =\nxs(4,4)$ and setting the excitation frequency to~$\omega\nes=\nes 5$. In this case, the wavelength-to-fracture-size  ratio can be computed as $\lambda_s/L \simeq 1.8$. The resulting distributions of~$\breve{\Tcal}$ and point-optimal normal~$\bn'$ are shown in Fig.~\ref{Lcurve}, from which one can observe that i) the method performs similarly for both planar an non-planar fractures, and ii) the reconstruction procedure is apparently not sensitive to the nature of the interfacial condition in terms of~$k_s$ and~$k_n$. 

\begin{figure}[!h]
\vspace*{2mm} 
\center\includegraphics[width=0.9\linewidth]{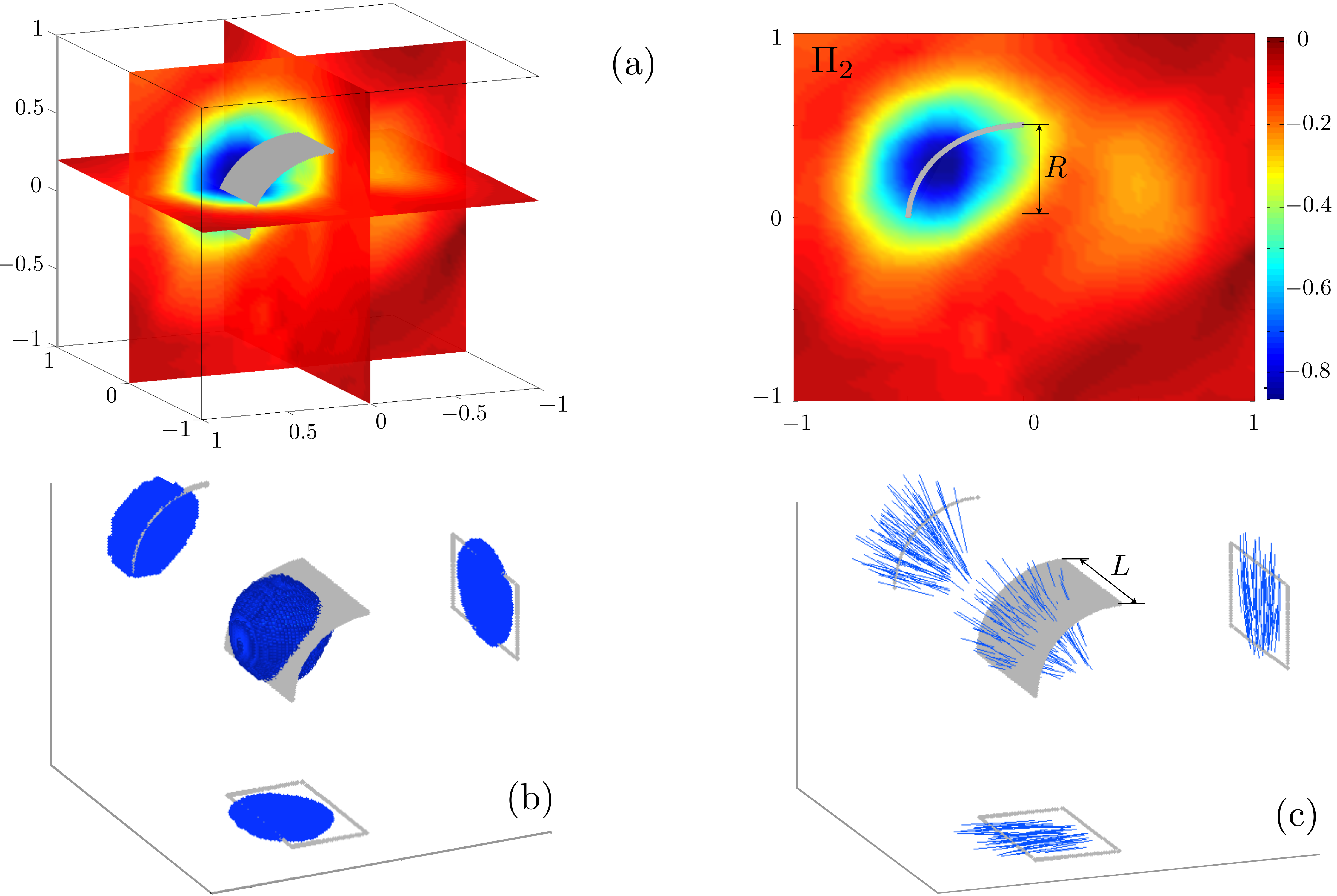} 
\caption{Low-frequency reconstruction of a cylindrical fracture with $(k_s,k_n)\nxs =\nxs(4,4)$: (a)~composite TS distribution $\breve{\Tcal}(\bxio;\bn',0,0)$ in three dimensions, and in the fracture mid-section $\Pi_2$; (b)~region containing the most pronounced negative TS values: $-1\nes\leqslant\nes\breve{\Tcal}\nes\leqslant\nes -0.6$, and (c)~point-optimal normal vector~$\bn'(\bxio)$ plotted over the true fracture surface $\Gamma$.} \lb{Lcurve}\vspace*{0.0mm}
\end{figure}

\paragraph*{Interface characterization} 

With reference to~ Fig.~\ref{Num_setup}(a), the characterization of a penny-shaped fracture with three distinct interface scenarios is considered, namely: i)~the traction-free crack i.e.~$k_n=k_s=0$, ii)~isolated fluid-filled fracture with~$(k_s,k_n)=(2,100)$, and  
iii)~fracture with rough surfaces under insignificant normal stress~\cite{Seidel1995}, simulated by setting~$(k_s,k_n) = (100,2)$. In all cases, the interface  characterization is carried out at $\omega =5$ i.e. $\lambda_s/L \simeq 1.8$ using twelve incident (P- and S-) waves as described earlier. Next, to identify the contact condition, the ``test'' indicator functions 
$$\breve{\Tcal}_1(\bxio) ~=~ |\breve{\Tcal}_\text{\nes m}|^{-1} \,  \frac{\sum_{\text{n}=1}^{12} \! \Tcal(\bxio;\bn,0,50)\big|_{(\bb_{\text{n}}, \bd_{\text{n}})}}{\sum_{\text{n}=1}^{12} \! |\sigma_{33}^f(\bz)|^2_{(\bb_{\text{n}}, \bd_{\text{n}})}}, \qquad \breve{\Tcal}_2(\bxio) ~=~  |\breve{\Tcal}_\text{\nes m}|^{-1} \,  \frac{\sum_{\text{n}=1}^{12} \! \Tcal(\bxio;\bn,50,0)\big|_{(\bb_{\text{n}}, \bd_{\text{n}})}}{\sum_{\text{n}=1}^{12} \nxs \big[\sigma_{3\beta}^{f*}  \sigma_{3\beta}^f \big] (\bz)\big|_{(\bb_{\text{n}}, \bd_{\text{n}})}}$$
are computed on the basis of~(\ref{Rec_sch}), where $\breve\Tcal_\text{\nes m} = \min \exs \lbrace  \text{min}_{\bxio} \breve{\Tcal}_1, \text{min}_{\bxio} \breve{\Tcal}_2 \rbrace $. 
The resulting distributions of~$\breve\Tcal_1$ and~$\breve\Tcal_2$ for all three scenarios are plotted in Fig.~\ref{ID}, using a common color scale, over the fracture's mid-section~$\Pi_1$. From the display, it is clear that $\breve{\Tcal}_1(\bxio)/\breve{\Tcal}_2(\bxio)\nes =\nes O(1)$, $\, \breve{\Tcal}_1(\bxio)\nes\ll\nes\breve{\Tcal}_2(\bxio)$, and $\breve{\Tcal}_1(\bxio)\nes\gg\nes\breve{\Tcal}_2(\bxio)$ respectively for the ``true'' interfacial conditions according to i), ii) and~iii). These results indeed support the claim of the preliminary characterization scheme that, at long illumination wavelengths, the interfacial condition of nearly-planar fractures can be \emph{qualitatively} assessed at virtually no computational cost - beyond what is needed to identify the fracture geometrically. 
\begin{figure}[Hbp]
\vspace*{0mm} 
\center\includegraphics[width=0.92\linewidth]{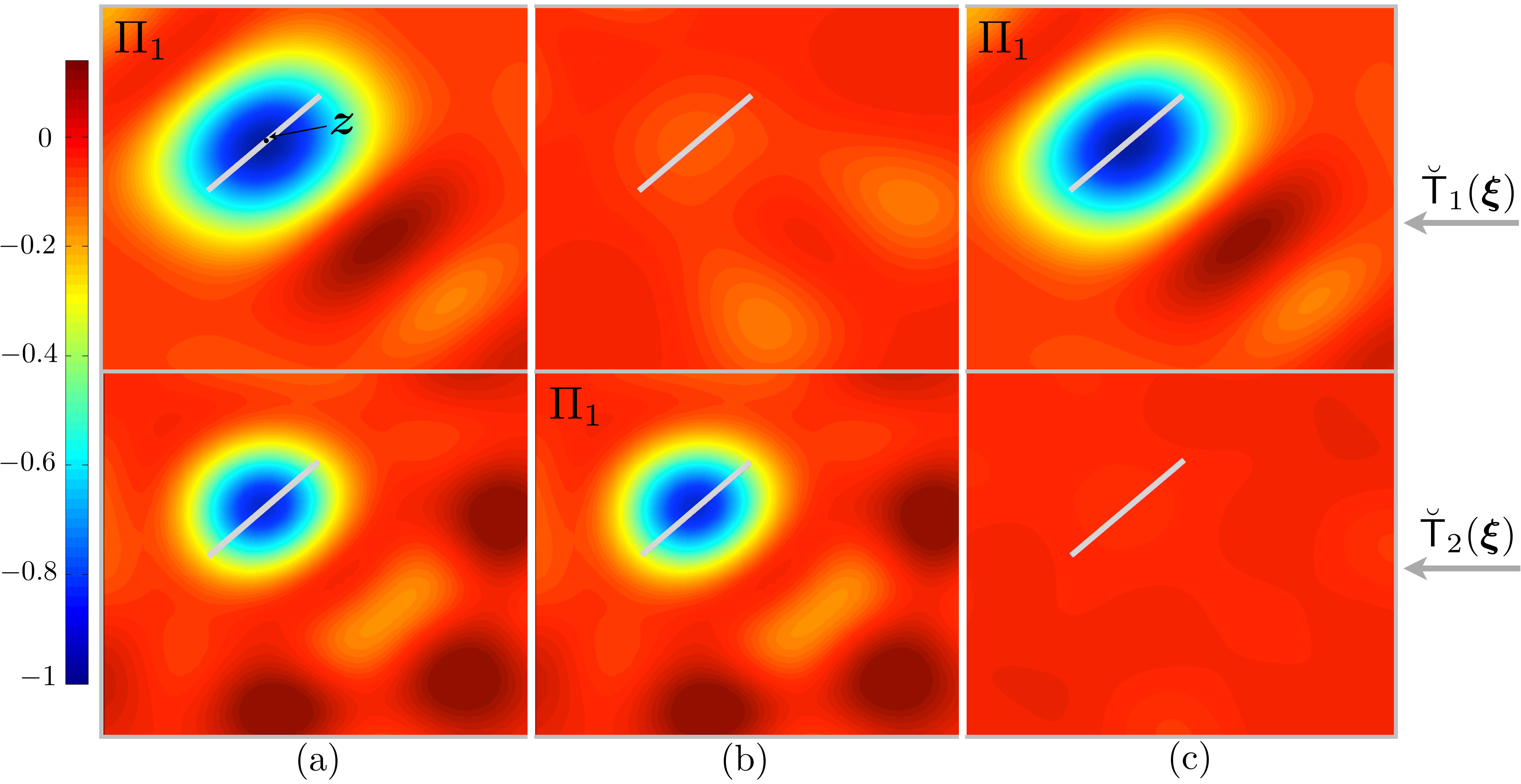} 
\caption{Spatial distribution of~$\breve{\Tcal}_1$~(top panels) versus~$\breve{\Tcal}_2$~(bottom panels), in the mid-section of a penny-shaped fracture, whose interface is (a) traction-free ($k_s=k_n =0)$, (b) isolated fluid-filled  $(k_s\nes=2, k_n\nes=100)$, and (c) of significant surface roughness and under low normal stress $(k_s\nes=100, k_n\nes=2)$.} \lb{ID}\vspace*{0.0mm}
\end{figure}

\paragraph*{TS reconstruction at higher frequencies} 

Clearly, the geometrical information in Figs.~\ref{LFN}(b) and~\ref{Lcurve}(b), obtained with \mbox{$\lambda_s/L\sim 2$}, does not carry sufficient detail to accurately reconstruct the fracture surface in~3D from the scattered elastic waves. To help mitigate the drawback, the forward scattering problem is recomputed at a higher frequency, namely $\omega = 20\,$ for which $\lambda_s/L \simeq 0.45$. In doing so, the number of incident elastodynamic fields is increased so that \emph{twenty} plane waves of each (P- and S-) type, propagating in directions $\lbrace \bd_{\text{n}}\!\in \Omega, \, \, \text{n}=1,2,...,20 \rbrace$ -- evenly distributed over the unit sphere~$\Omega$ -- participate in the TS evaluation. The resulting ``high-frequency'' behavior of the composite indicator function
\[
\breve{\Tcal}(\bxio;\bn',0,0) ~=~ |\text{min}_{\bxio}\breve{\Tcal}|^{-1} \, \sum_{\text{n}=1}^{40} \Tcal(\bxio;\bn',0,0)|_{(\bb_{\text{n}}, \bd_{\text{n}})} 
\]
is shown in Figs.~\ref{HF} and~\ref{curved} for the cases of penny-shaped fracture and cylindrical fracture, respectively (see Fig.~\ref{Num_setup}). The featured results are consistent with the recent findings in acoustics and elastodynamics~\cite{Fei2004,Tokm2013,Fatemeh2014} which demonstrate that, at higher illumination frequencies, pronounced negative values of TS 
\begin{figure}[h]
\vspace*{0mm} 
\center\includegraphics[width=0.82\linewidth]{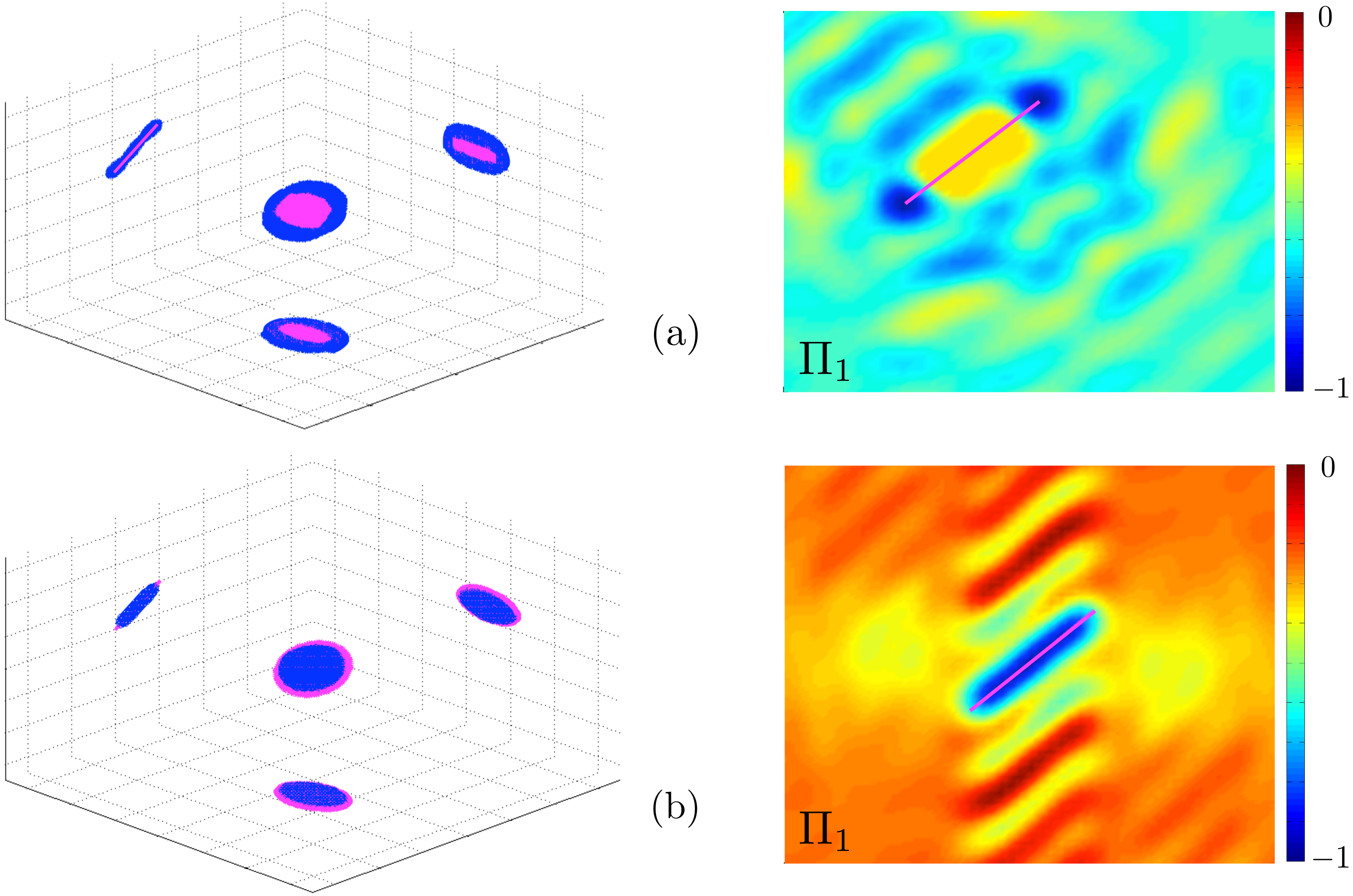} 
\caption{``High''-frequency ($\lambda_s/L \simeq 0.45$) TS reconstruction of a penny-shaped fracture whose interfacial condition is~(a) traction-free ($k_s=k_n=0$), and (b) isolated fluid-filled $(k_s=2,k_n=100)$.} \lb{HF}\vspace*{0.0mm}
\end{figure} 
tend to localize in a narrow region \emph{tracing the boundary} of a scatterer. However, it the present case~$\breve\Tcal$ also exhibits a notable sensitivity to the fracture's interfacial condition; in particular, for the traction-free crack in Fig.~\ref{HF}(a), extreme negative values of the TS are localized in the vicinity of the \emph{crack tip}, whereas in the case of fractures with interfacial stiffness -- Fig.~\ref{HF}(b) and Fig.~\ref{curved} --
the TS exposes the \emph{entire fracture surface}. These initial results suggest that at shorter wavelengths, the TS experiences a different type of sensitivity to the fracture interfacial condition, which may lead to a more detailed identification of the fracture's specific stiffnesses. Given a sensory data set that includes the scattered field measurements at both long and short wavelengths, one may consider a staggered approach where i)~high-frequency data are deployed to more precisely evaluate the fracture geometry, including its location~$\bz$ and effective normal~$\bn$; $\,$ii) low-frequency observations are used as in Section~\ref{sec4} to qualitatively identify the interfacial condition, and iii) additional information is obtained on the fracture's interface thanks to the dependence of the high-frequency TS thereon. Such developments, however, require high-frequency asymptotics of the scattered field due to a fracture with specific stiffness -- a topic that is beyond the scope of this study.

\begin{figure}[ht]
\vspace*{0mm} 
\center\includegraphics[width=0.96\linewidth]{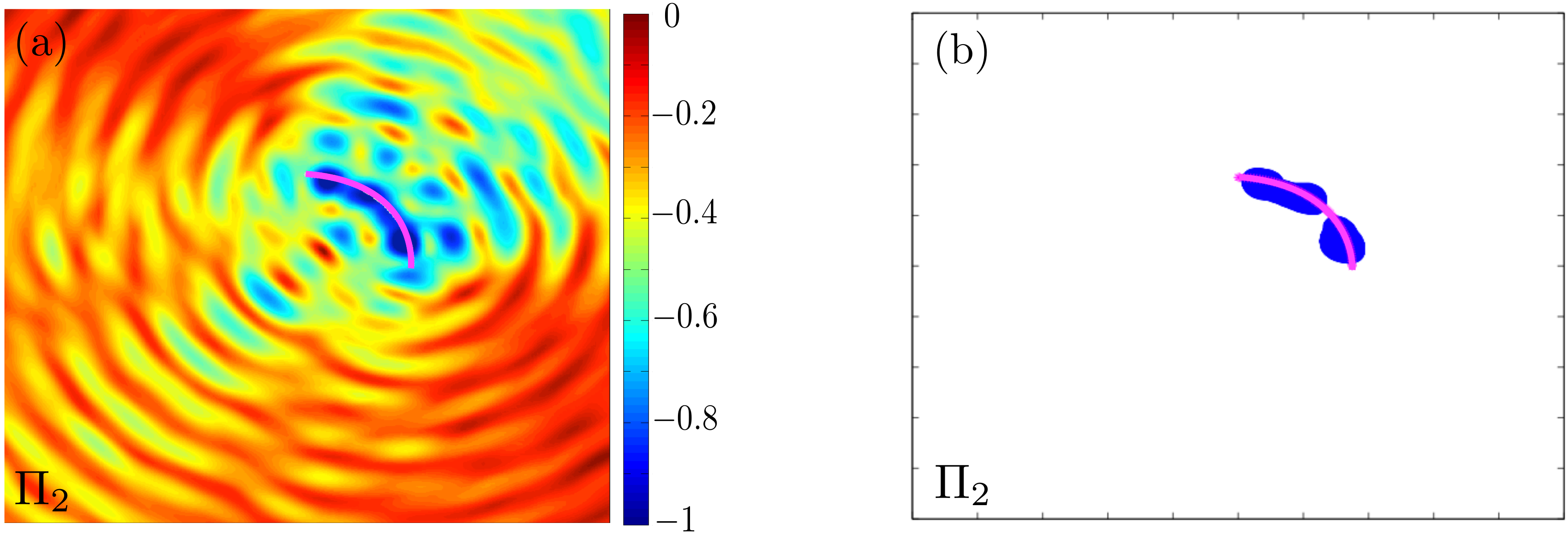} 
\caption{``High''-frequency ($\lambda_s/L \simeq 0.45$) TS reconstruction of a cylindrical fracture with $k_n=k_s=4$: (a) distribution of~$\breve\Tcal$ in the fracture's mid-section, and (b) corresponding $\breve\Tcal$~map thresholded at~70\%.} \lb{curved}\vspace*{0.0mm}
\end{figure}

\section{Summary} 

\noindent This work investigates the utility of the topological sensitivity (TS) approach as a non-iterative tool for the waveform tomography of fissures with specific stiffness, e.g.~hydraulic fractures. On postulating the nucleation of an infinitesimal  penny-shaped fracture with constant (normal and shear) interfacial stiffnesses at a sampling point, the TS formula and affiliated elastic polarization tensor are calculated and expressed in closed form. In this setting, it is shown via numerical simulations that the TS carries the capacity of exposing the \emph{fracture location} and its \emph{unit normal} from the long-wavelength scattered field measurements, regardless of the assumption on the (trial) interfacial parameters of a vanishing fracture. Given thus obtained geometric information and elastodynamic (free- and adjoint-field) simulations required for its computation, it is further shown that the interfacial condition of nearly-planar fractures can be \emph{qualitatively} identified at virtually no added computational cost, using two auxiliary TS maps evaluated for certain (extreme) combinations of the trial contact parameters. In particular, the analysis shows that such scheme allows for \emph{the ratio} between the shear and normal specific stiffness -- representative of a hidden fracture -- to be exposed as either i)~near-zero, ii) on the order of unity, or iii) exceeding unity by a large amount. Such information can be used both directly, e.g.~to discriminate between the old, newly created and proppant-filled fractures, and as an ``initial guess'' on the fracture's interfacial condition for a more comprehensive, nonlinear optimization approach to waveform tomography. Through preliminary simulations at ``short'' incident wavelengths -- subpar to the characteristic fracture size -- which demonstrate both enhanced imaging resolution of the TS indicator function and its heightened sensitivity to fracture's interfacial condition, this study further provides an impetus for studying the \emph{high-frequency} elastodynamic scattering by, and TS sensing of, fractures with specific stiffness.      

\section*{Acknowledgment} 

\noindent The support provided by the U.S. Department of Energy via NEUP Grant \#10-862 and the University of Minnesota Supercomputing institute is kindly acknowledged.

\appendix

\section{Asymptotic behavior of the integrals in~(\ref{mathC})}  \label{App-A}

\noindent The aim of this section is to expose the leading-order behavior of the integrals in~(\ref{mathC}) for $R_2/R_1\nes\gg\nes 1$, where the integrands are expressed as antilinear forms in terms of the components of the fundamental stress tensor.    

\paragraph*{Elastodynamic fundamental solution} 

Assuming time-harmonic excitation by the unit point force applied at $\bx\in\mathbb{R}^3$ in direction~$k$, the governing equation of motion for an infinite solid with shear modulus~$\mu$, mass density~$\rho$ and Poisson's ratio~$\nu$ can be written as 
\[
\nabla_{\bxi} \sip \hs \bSig^k \hs  + \rho \exs \omega^2 \exs {\bU^k} \hs=\hs \delta(\bxi-\bx) \be_k, \qquad \bxi\in \mathbb{R}^3, \quad \bxi\neq\bx,
\]
where $\bU^k(\bxi,\bx,\omega)$ and $\bSig^k(\bxi,\bx,\omega)$ denote respectively the fundamental displacement vector and stress tensor, given in the component form by 
\beq
\begin{aligned}
\lb{HFS}  
 U_i^k(\bxi,\bx,\omega) &~=~ \frac{1}{4\pi \mu r} \big( B_1(r) \exs \delta_{ik} + B_2(r) \exs  r_{,i}  r_{,k} \big),  \\*[0.5mm] 
\Sigma_{ij}^k(\bxi,\bx,\omega) &~=~  \frac{1}{4\pi r^2} \big( 2 \exs B_3(r) \exs r_{,i} r_{,j} r_{,k}  +  (\delta_{ik} \exs r_{,j}+\delta_{jk} \exs r_{,i}) B_4(r) + B_5(r) \exs \delta_{ij} r_{,k} \big).
\end{aligned} 
\eeq 
Here
\beq
\begin{aligned}
\lb{Cons}  
B_1(r) &~=~ \Big( 1-\frac{\text{i}}{\chi_s}-\frac{1}{{\chi_s}^2}\Big) e^{-\text{i}\chi_s} + \gamma^2 \Big( \frac{\text{i}}{\chi_p}+\frac{1}{{\chi_p}^2}\Big) e^{-\text{i}\chi_p}, \\*[0.5mm] 
B_2(r) &~=~ \Big( \frac{3}{{\chi_s}^2}+\frac{\text{3i}}{\chi_s}-1\Big) e^{-\text{i}\chi_s} - \gamma^2 \Big(\frac{3}{{\chi_p}^2}+ \frac{\text{3i}}{\chi_p}-1\Big) e^{-\text{i}\chi_p}, \\*[0.5mm]
B_3(r) &~=~ \Big(6- \frac{15}{{\chi_s}^2}-\frac{\text{15i}}{\chi_s}+\text{i}\chi_s \Big) e^{-\text{i}\chi_s} - \gamma^2 \Big(6- \frac{15}{{\chi_p}^2}-\frac{\text{15i}}{\chi_p}+\text{i}\chi_p \Big) e^{-\text{i}\chi_p}, \\*[0.75mm]
B_4(r) &~=~ -(1+\text{i}\chi_s)e^{-\text{i}\chi_s}+2 B_2(r),\\*[1.1mm]
B_5(r) &~=~ -(1-2\gamma^2)(1+\text{i}\chi_p)e^{-\text{i}\chi_p}+2 B_2(r),
\end{aligned} 
\eeq 
and   
\[
r = |\bxi-\bx|, \quad \chi_s = \frac{r\omega}{c_s}, \quad \chi_p = \frac{r\omega}{c_p}, \quad \gamma = \frac{c_s}{c_p}, \quad r_{,i} = \frac{\partial r}{\partial \xi_i}, 
\]
where $\delta_{ij}$ denotes the Kronecker delta, $c_s=\sqrt{\mu/\rho}$, and~$c_p=c_s\sqrt{(2\nes-\nes2\nu)/(1\nes-\nes2\nu)}$. 

\paragraph*{Integration} 

For $\bx\nes\in\nes\mathcal{B}_1$ and~$\bxi\nes\in\nes\So$ (see Fig.~\ref{setup}), the argument $r = |\bxi-\bz|$ of the relevant coefficients in~(\ref{HFS}) and (\ref{Cons}) describing $\Sigma^k_{ij}(\bxi,\bz,\omega)$ can be approximated (assuming $R_2/R_1\nes\gg\nes 1$) to the leading order as $r\simeq R_2$; as a result, the prevailing behavior of integrals in~(\ref{mathC}) can be written as 
\beq 
\lb{C1} Q_1: \quad
\int_{\So}  [\Sigma^{j*}_{33} \exs  \Sigma^j_{33}](\bxi,\bz,\omega) \, \text{d}S_{\bxi} ~\simeq~ \text{X}_1  \int_{0}^{\pi} \!\! \int_{0}^{2\pi}  \! \big( \text{X}_2 \exs r_{\!,3}^4+ 4\exs \text{Re}(\text{X}_3) \exs r_{\!,3}^2+ |B_5(R_2)|^2  \big) \nxs \sin(\phi) \exs \text{d}\theta \exs \text{d}\phi, 
\eeq
\beq
\lb{C_mixed}  Q_2: \quad
\int_{\So}   [\Sigma^{j*}_{33} \exs  \Sigma^j_{3\beta}](\bxi,\bz,\omega) \, \text{d}S_{\bxi} ~\simeq~ \text{X}_1  \int_{0}^{\pi} \!\! \int_{0}^{2\pi}  r_{\!,3} \exs {r_{\!,\beta}} \exs \big( \text{X}_2 \exs r_{\!,3}^2+2\exs \text{X}_3  \big) \nxs \sin(\phi) \exs \text{d}\theta \exs \text{d}\phi \,=\,0,
\eeq
\beq
\lb{C3}  Q3: \quad
\begin{aligned}
&\int_{\So}   | \Sigma^j_{3\beta}(\bxi,\bz,\omega)|^2 \, \text{d}S_{\bxi} ~\simeq~ \text{X}_1  \int_{0}^{\pi} \!\! \int_{0}^{2\pi}  \! \big( \text{X}_2 \exs r_{\!,\beta}^2 \exs r_{\!,3}^2+  \exs (r_{\!,3}^2 +r_{\!,\beta}^2) |B_4(R_2)|^2 \big) \nxs \sin(\phi) \exs \text{d}\theta \exs \text{d}\phi,  \\*[1.0mm] 
&\int_{\So}   [\Sigma^{j*}_{31} \exs  \Sigma^j_{32}](\bxi,\bz,\omega) \, \text{d}S_{\bxi} ~\simeq~ \text{X}_1  \int_{0}^{\pi} \!\! \int_{0}^{2\pi}  \! \big( \text{X}_2 \exs {r_{\!,1}} \exs {r_{\!,2}} \exs r_{\!,3}^2+ |B_4(R_2)|^2 \exs{r_{\!,1}}{r_{\!,2}} \big) \nxs \sin(\phi) \exs \text{d}\theta \exs \text{d}\phi \,=\,0, 
\end{aligned}
\eeq
where $\beta = 1,2\,$ and 
\beq
\lb{X}
\text{X}_1 = \frac{1}{(4 \exs \pi R_2)^2}, \quad \text{X}_2 = 4 \text{Re} \big\lbrace \big[ \exs |B_3|^2+2 B_3^* \nxs B_4 \big] \nxs (R_2) \big\rbrace, \quad \text{X}_3 =\big[ |B_4|^2+ (B_3+B_4) B_5^* \big] \nxs (R_2).
\eeq
Considering the unit vector~$\nabla r = \big(r_{\!,1},r_{\!,2},r_{\!,3}\big) = \big(\sin(\phi)\cos(\theta),\sin(\phi)\sin(\theta),\cos(\phi)\big)$ used to define $\So$ in spherical coordinates, the integrals of~(\ref{C1})-(\ref{C3}) are analytically evaluated which results in~(\ref{TS_str2}).   

\section{BIE computational platform}  \label{App-B}

\noindent This section summarizes the numerical scheme adopted to solve the elastodynamic traction BIE for a fractured three-dimensional solid. The approach borrows substantially from the ideas established in~\cite{Bon1999,Bojan1996} considering the regularization of the featured surface integrals. For brevity, the technique is described with reference to the elastostatic canonical problem~(\ref{canonic}). With slight modifications, however, this method is utilized in a more general setting of Section~{\ref{sec5}} to compute the scattering of elastic waves by an \emph{arbitrarily-shaped} fracture. Accordingly, the auxiliary formulae are expressed in their most general form (as applicable). Unless stated otherwise, the Einstein summation convention is assumed over repeated coordinate indexes.       

\paragraph*{Regularization} 

To avoid evaluating the Cauchy principal value in~(\ref{canonic}), the featured integral equation can be conveniently rewritten as 
\beq
\begin{aligned}
\lb{RTBIE}  
&\frac{1}{2} \exs {[ \be_i \otimes \be_j +\be_j \otimes \be_i ]}_{k\ell} \exs n'_\ell\,-\, {[K_{\mbox{\tiny{trial}}}]}_{k\ell} \exs \llbracket V_\ell \rrbracket^{ij}(\bar\bxi) ~=~ n'_\ell \exs C_{k\ell pq} \exs D_{ \nxs qs} \llbracket V_m \rrbracket^{ij} \nxs (\bar\bxi) \exs \check{S}^{ \exs p}_{ms} (\bar\bxi, \Ga) \,+ \\*[1mm] 
& \qquad \qquad  n'_\ell \exs C_{k\ell pq}  \int_{\Ga} \! \big(  D_{qs}  \llbracket V_m \rrbracket^{ij}  (\bar\bx) -  D_{qs}  \llbracket V_m \rrbracket^{ij}  (\bar\bxi)  \big) \exs \check\Sigma_{ms}^{p}(\bar\bxi,\bar\bx)  \, \text{d}S_{\bar\bx},  \qquad    \bar\bxi \in \Ga,
\end{aligned} 
\eeq 
where the dummy indexes are summed over~$\overline{1,3}$, and the singularity is transferred to the auxiliary integrals 
\beq
\lb{KSint}
I_\ell(\bar\bxi, \Ga) ~=~ \dashint_{\Ga} \frac{1}{\text{r}^2} \exs \text{r}_{,\ell} \,\, \text{d}S_{\bar\bx}, \qquad J_{ pq \ell}(\bar\bxi, \Ga) ~=~ \dashint_{\Ga} \frac{1}{\text{r}^2} \exs \text{r}_{,p} \exs \text{r}_{,q} \exs \text{r}_{,\ell} \,\, \text{d}S_{\bar\bx}, \qquad    \bar\bxi \in \Ga, 
\eeq
comprising the third-order tensor   
\beq
\lb{KSing}
\check{S}^{ \exs p}_{ms}(\bar\bxi, \Ga)~=~ \dashint_{\Ga} \check\Sigma^{p}_{ms} (\bar\bxi, \bar\bx) \exs \text{d} S_{\bar\bx}~=~-\frac{1}{8\pi(1-\nu)} \big[ (1-2\nu) (\delta_{mp} \exs I_s + \delta_{sp} \exs I_m - \delta_{ms} \exs I_p) + 3 J_{msp} \big].
\eeq
Here it is useful to recall that $\text{r} = | \bar\bxi - \bar\bx|$, $\, \text{r}_{,\ell} = \partial \text{r}/\partial \bar{x}_\ell$, and $\, \bK_{\mbox{\tiny{trial}}}=\kappa_n (\bar\be_3 \otimes \bar\be_3)+\kappa_s (\bar\be_1 \otimes \bar\be_1)+\kappa_s (\bar\be_2 \otimes \bar\be_2)$ where $(\bar\be_1,\bar\be_2,\bar\be_3)$ are the unit vectors along $(\bar\xi_1,\bar\xi_2,\bar\xi_3)$. Considering~$I_\ell$ first, one finds~\cite{Bon1999}  via integration by parts, taking advantage of the Stokes identity, and noting that  $\dbbV^{ij}\nes=\nes \boldsymbol{0}$ on~$\partial\Ga$, that the first in~(\ref{KSint}) can be reduced as
\beq
\lb{Il}  
I_\ell~=~\int_{\Ga} \frac{1}{\text{r}} \exs \big( \frac{1}{\text{r}} \exs n'_p \exs \text{r}_{,p} - w_{q}  n'_q\big) n'_{\ell} \,\, \text{d}S_{\bar\bx} - \int_{\partial \Ga} \! \frac{1}{\text{r}} \exs  \text{v}_{\ell} \,\, \text{d}s, \qquad \bar\bxi\in\Ga,
\eeq
where $\Ga$ is interpreted as an \emph{open set} (excluding fracture's edge $\partial\Ga$);  $p,q=1,2,3$; $\textbf{v}$ denotes the outward normal on~$\partial \Ga$ lying within the tangent plane to $\Ga$, and $w_k(f) = f_{,k} - n'_k (f_{,p} n'_p)$ is the tangential derivative operator. In terms of $J_{pq\ell}$, on the other hand, one similarly obtains 
\beq
\begin{aligned}
\lb{Jpql}  
3 J_{ pq \ell}~=~\delta_{p\ell} \exs I_{q} +  \delta_{q\ell}  \exs I_{p} \,+ &  \int_{\Ga} \frac{1}{\text{r}} \big( \text{r}_{,q} \exs \text{r}_{,s} \exs D_{sp} n'_{\ell} -w_{p} (n'_q n'_\ell) -w_{q} (n'_p n'_\ell) \big) \,\, \text{d}S_{\bar\bx} \,+ \\*[1.2mm] 
& \delta_{pq} \int_{\Ga} \frac{1}{\text{r}^2} \exs n'_{\ell} \exs n'_{s} \exs  \text{r}_{,s} \,\, \text{d}S_{\bar\bx} + \int_{\Ga} \frac{1}{\text{r}} \exs n'_{\ell} (2n'_{p} \exs n'_{q}-\text{r}_{,p} \exs \text{r}_{,q}) w_{s}  n'_{s} \,\, \text{d}S_{\bar\bx} \, - \\*[1.2mm] 
& 2 \int_{\Ga} \frac{n'_{p}}{\text{r}^2} \exs n'_{q} \exs n'_{\ell} \exs n'_{s} \exs \text{r}_{,s} \,\, \text{d} S_{\bar\bx} + \int_{\partial \Ga} \! \frac{1}{\text{r}} \big( \text{v}_{p} \exs n'_{\ell} \exs n'_{q} + \text{v}_{q} \exs n'_{\ell} \exs n'_{p} - \text{r}_{,p} \exs \text{r}_{,q} \exs \text{v}_{\ell} \big) \,\, \text{d}s \, +  \\*[1.2mm] 
&\int_{\partial \Ga} \! \frac{1}{\text{r}} \exs n'_{\ell} \exs \text{r}_{,q} \text{r}_{,s} \big( \text{v}_s n'_{p} - \text{v}_p n'_{s} \big) \,\, \text{d}s, \qquad \qquad \quad \,\,\, \bar\bxi\in\Ga.
\end{aligned} 
\eeq
For further details on~(\ref{RTBIE})--(\ref{Jpql}), the reader is referred to Chapter~13 in~\cite{Bon1999}. Here one should mention that, for the canonical problem in Section~\ref{PLT} where~$\bn' = \bar\be_3$, formulas~(\ref{Il}) and~(\ref{Jpql}) can be remarkably simplified (see Appendix 5.A in~\cite{Bon1999}). In this presentation, however, both formulae are kept in their general format for they also pertain to the scattering by arbitrarily-shaped fractures. 

\paragraph*{Parametrization} 

With reference to Fig.~\ref{mesh}, the fracture boundary $\Ga$ is discretized using a conformal mesh permitting surface parametrization ($\by \rightarrow \bar\bx$) as   
\beq
\lb{Gpara}
\bar\bx(y_1,y_2) ~=~\psi_{\text{m}} (y_1,y_2) \,\, \bar\bx^{\text{m}}, \qquad \text{m}~=~1, \cdot \! \cdot \! \cdot ,\text{N}_n, \qquad  -1\leqslant y_1, y_2 \leqslant 1.
\eeq    

\begin{figure}[!h]
\vspace*{0mm} 
\center\includegraphics[width=1\linewidth]{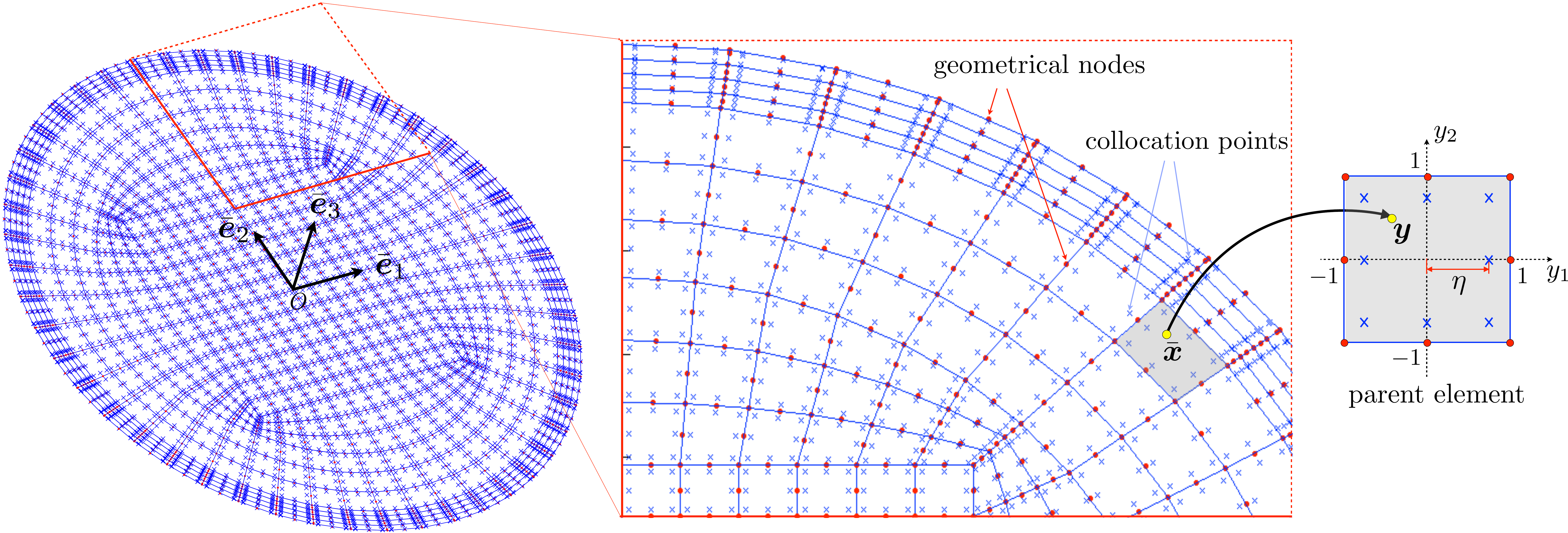} 
\caption{Geometric (dotes) and interpolation (crosses) nodes on $\Ga$ in physical and parametric spaces.} \lb{mesh}\vspace*{0.0mm}
\end{figure} 
\noindent
Here $\text{N}_n =8$ designates the number of nodes per element, and $\bar\bx^{\text{m}}$ denotes the global coordinates of the element's $\text{m}^{\text{th}}\!$ node -- whose shape function $\psi_{\text{m}} (\by)$ is that of the standard eight-node quadrilateral element. In this setting, one finds the natural basis $\ba_{1,2}$ of the tangent plane and the surface differential as   
\beq
\lb{DG}
\ba_\beta(\by) ~=~ \frac{\partial\psi_{\text{m}}}{\partial y_\beta}  \, \bar\bx^{\text{m}},  \qquad \text{d}S_{\bar\bx}~=~ \mathcal{G}(\by) \exs \text{d}S_{\by}, \qquad \mathcal{G}(\by)~=~ | \ba_1 \! \times \! \exs \ba_2 |,
\eeq  
where $\beta\nes=\nes1,2$ and the dummy index m is summed over~$\overline{1,8}$. At this stage, one should note that all integrands in~(\ref{Il}) and~(\ref{Jpql}) -- comprising~$\check{S}_{ms}^p$ in~(\ref{RTBIE}) -- are \emph{known} so that the boundary parametrization, given by~(\ref{Gpara}) and~(\ref{DG}), is the only necessary step prior to numerical integration.    

In light of the smoothness requirement by the traction BIE~(\ref{RTBIE}) and the adverse presence of the tangential derivative operator $D_{qs} (\cdot)$, the COD $\dbbV^{ij}(\bar\bx)$ is discretized via \emph{non-conforming} interpolation (see Section 3.2 of~\cite{Bon1999} for details). In particular, the interpolation i.e.~collocation points are situated \emph{inside} the boundary elements (see Fig.~\ref{mesh}), and their position with respect to the geometrical nodes -- in each element -- is quantified via parameter $\eta$ in the $(y_1,y_2)$ space. In this setting, the COD over the parent element can be approximated as 
\beq
\lb{Ipara}
\llbracket V_{\ell} \rrbracket^{ij}(\bar\bx) ~=~\phi_{\text{m}} (\by) \, \llbracket V_\ell \rrbracket^{ij}_{\text{m}}, \qquad  \text{m}~=~1, \cdot \! \cdot \! \cdot ,8,
\eeq    
where
\beq
\lb{Ipara2}
\llbracket V_\ell \rrbracket^{ij}_{\text{m}}~=~\llbracket V_{\ell} \rrbracket^{ij}(\bar\bx^{\text{m}}), \qquad \phi_{\text{m}}(\by) ~=~ \frac{1}{4\eta^3}(\eta+ y_1^{\text{m}}  y_1)(\eta+ y_2^{\text{m}}  y_2)(y_1^{\text{m}}  y_1+ y_2^{\text{m}}  y_2-\eta),
\eeq   
and $(y_1^{\text{m}},y_2^{\text{m}})$ is the position of the m$^\text{th}\!$ collocation point in the parent element as shown in Fig.~\ref{mesh}. Here it is important to mention that the surface elements adjacent to~$\partial \Ga$ are of \emph{quarter-node type} (see e.g.~Chapter 13 of~\cite{Bon1999}), designed to reproduce the square-root behavior of the COD in the  vicinity of~$\partial \Ga$. Note that for constant distribution of interfacial stiffness, i.e.~for constant stiffness matrix $\bK_{\mbox{\tiny{trial}}}$, the asymptotic behavior of the COD near $\partial \Ga$ remains the same as that in the case of traction-free crack (see~\cite{Ueda2006} for proof). 

Given~(\ref{Gpara})--(\ref{Ipara2}), it can be shown~\cite{Bon1999} that the tangential derivative operator $D_{qs} \llbracket V_\ell \rrbracket^{ij}(\bar\bx)$ in~(\ref{RTBIE}) permits the parameterization 
\beq
\lb{Dqs}
D_{qs} \llbracket V_{\ell} \rrbracket^{ij}(\bar\bx)= \Lambda_{qs}^{\!\text{m}} (\by) \exs \llbracket V_\ell \rrbracket^{ij}_{\text{m}}, \quad \Lambda_{qs}^{\!\text{m}} (\by) = \frac{\epsilon_{pqs}}{\mathcal{G} \nxs (\by)} \exs \big{[} (\ba_2 \sip \bar\be_p ) \phi_{\text{m},1} -(\ba_1 \sip \bar\be_p ) \phi_{\text{m},2} \big{]} \nxs (\by) , \quad \phi_{\text{m},\beta}=\frac{\partial\phi_\text{m}}{\partial y_\beta},
\eeq   
where the dummy indexes $p$ and m are summed over $\overline{1,3}$, and $\overline{1,8}$, respectively; the basis unit vectors $\bar\be_p$ are shown in Fig.~\ref{mesh}, and $\epsilon_{pqs}$ denotes the Levi-Civita symbol.    

On substituting~(\ref{DG}) and~(\ref{Dqs}) into~(\ref{RTBIE}), one arrives at the algebraic system for the values of~$\dbbV^{ij}$ at the collocation nodes  $\bar\bxi = \bar\bx^{ \text{m}^{\nxs*}}_{ e^{\nxs*}}\!$ as 
\beq
\begin{aligned}
\lb{TBEM}  
&\frac{1}{2} \exs {[ \be_i \otimes \be_j +\be_j \otimes \be_i ]}_{k\ell} \exs n'_\ell\,-\, {[K_{\mbox{\tiny{trial}}}]}_{k\ell} \exs  \llbracket V_\ell \rrbracket^{ij}_{\text{m}^{\nxs*} \nxs e^{\nxs*}} ~=~ n'_\ell \exs C_{k\ell pq} \exs \Lambda_{qs}^{\!\text{m} e^{\nxs*}} \! \nxs (\by^{\text{m}^{\nxs *}}\nxs)  \check{S}^{ \exs p}_{ms} (\bar\bx^{ \text{m}^{\nxs*}}_{ e^{\nxs*}}\!\!, \Ga) \llbracket V_m \rrbracket^{ij}_{\text{m} e^{\nxs*}}   \,+ \\*[-0.5mm] 
& \quad\, \sum_{e=1}^{\text{N}_e} \exs  \sum_{\text{m}=1}^{\text{N}_n} \Big\lbrace  n'_\ell \exs C_{k\ell pq}  \int_{S_{\by}} \! \big( \Lambda_{qs}^{\!\text{m} e} \nxs (\by) - \delta_{ee^{\nxs *}} \exs  \Lambda_{qs}^{\!\text{m} e} \nxs (\by^{\text{m}^{\nxs *}}\nxs)   \big) \exs \check\Sigma_{ms}^{p} \big{(}\bar\bx^{ \text{m}^{\nxs*}}_{ e^{\nxs*}}\!,\bar\bx(\by) \big{)} \mathcal{G}_e(\by)  \, \text{d}S_{\by} \Big{\rbrace} \llbracket V_m \rrbracket^{ij}_{\text{m} e}  \,- \\*[-0.5mm]
& \quad\,  \sum_{e=1}^{\text{N}_e} (1-\delta_{ee^{\nxs *}}) \!\!  \sum_{\text{m}=1}^{\text{N}_n} \Big\lbrace  n'_\ell \exs C_{k\ell pq} \Big[  \int_{S_{\by}} \!  \check\Sigma_{ms}^{p} \big{(}\bar\bx^{ \text{m}^{\nxs*}}_{ e^{\nxs*}}\!,\bar\bx(\by) \big{)} \mathcal{G}_e(\by)  \, \text{d}S_{\by} \Big]  \Lambda_{qs}^{\!\text{m} e^{\nxs *}} \! (\by^{\text{m}^{\nxs *}}\!) \Big{\rbrace} \llbracket V_m \rrbracket^{ij}_{\text{m} e^{\nxs*}} \,, 
\end{aligned} 
\eeq   
where $e^{\nxs*}$ is the element number; $\text{m}^{\nxs*}$ denotes the local node number; no sum over $e^{\nxs *}$ and~$\text{m}^{\nxs *}$ is implied; $N_n=8$;  $N_e$ is the number of elements, and
\beq
\lb{TBEM2}
\llbracket V_\ell \rrbracket^{ij}_{\text{m} e} ~=~\llbracket V_{\ell} \rrbracket^{ij}(\bar\bx^{ \text{m}}_e), \quad  \Lambda_{qs}^{\!\text{m} e} \nxs (\by)~=~ \frac{\epsilon_{pqs}}{\mathcal{G}_e \nxs (\by)} \exs \big{[} (\ba_2^e \sip \bar\be_p ) \phi_{\text{m},1} -(\ba_1^e \sip \bar\be_p ) \phi_{\text{m},2} \big{]} \nxs (\by). 
\eeq
Here it is worth noting that all integrals in~(\ref{Il}), (\ref{Jpql}) and~(\ref{TBEM}) are numerically integrable. A specific mapping technique (in the case of weak singularity) along with the standard Gaussian quadrature method is employed to evaluate the aforementioned integrals (see Section 3.9 of~\cite{Bojan1996} for details).

\end{document}